\documentclass[14pt]{article}
\pdfoutput=1
\usepackage{amsmath,amssymb,amsfonts,amsthm,bm,bbm,cancel,wasysym}
\usepackage{epsfig,graphics,graphicx,epstopdf}
\usepackage{array,booktabs,colortbl,colordvi,multirow}
\usepackage{colordvi,color,xcolor}
\usepackage{hyperref}
\usepackage{rotating}

\usepackage{verbatim}
\usepackage{cite}
\usepackage{subfig}
\usepackage{setspace}
\usepackage{url}
\usepackage[percent]{overpic}
\usepackage{slashed}
\usepackage{authblk}
\usepackage{xspace}
\usepackage{fullpage}
\usepackage{hyperref}

\newcommand{\unit}[1]{\ensuremath{\mathrm{\,#1}}\xspace}

\renewcommand{\photon}{\unit{photons}}

\def\ie{{\it i.e.}}
\def\eg{{\it e.g.}}
\def\etc{{\it etc}}

\def\to{\rightarrow}

\newskip\zatskip \zatskip=0pt plus0pt minus0pt
\def\matth{\mathsurround=0pt}
\def\lsim{\mathrel{\mathpalette\atversim<}}
\def\gsim{\mathrel{\mathpalette\atversim>}}
\def\atversim#1#2{\lower0.7ex\vbox{\baselineskip\zatskip\lineskip\zatskip
  \lineskiplimit 0pt\ialign{$\matth#1\hfil##\hfil$\crcr#2\crcr\sim\crcr}}}

\begin{document}


\begin{flushright}
SLAC-PUB-260112\\
\today
\end{flushright}
\vspace*{5mm}

\renewcommand{\thefootnote}{\fnsymbol{footnote}}
\setcounter{footnote}{1}

\begin{center}

{\Large {\bf A Model for Dark Moments of the $W$ Boson}}\\

\vspace*{0.75cm}

{\bf Thomas G. Rizzo}~\footnote{rizzo@slac.stanford.edu}

\vspace{0.5cm}

{SLAC National Accelerator Laboratory}\\ 
{2575 Sand Hill Rd., Menlo Park, CA, 94025 USA}

\end{center}
\vspace{.5cm}


\begin{abstract}
\noindent  
Loops of portal matter (PM) fields carrying both dark and Standard Model (SM) quantum numbers can lead to the kinetic mixing (KM) of the SM photon and the analogous dark photon (DP) of a 
phenomenologically interesting magnitude. However, in specific frameworks, different loops of these same PM fields can also lead to other new types of interactions between some of the 
SM fields and the DP via the generation of `dark moment'-like couplings, even though the SM fields carry a zero dark charge at tree-level. In recent work, this possibility has been 
explored for the case when these SM fields are fermionic. In this paper, we extend this idea to the case of the SM $W$ boson employing a previously examined model wherein PM consists of 
a complex scalar triplet plus a complex singlet, both of which obtain vevs and thus also generate the DP mass. While this setup results in a somewhat stronger interaction between the DP and the 
$W$ than in the familiar KM setup, the rate for $W+$DP production at the LHC is shown to still be rather small and is very difficult to observe due to large SM backgrounds.  The direct production of 
the scalar PM itself, however, is shown to lead to similar new physics signatures but with much larger rates and are considered here in their own right. The properties of such new PM states may 
already be constrained by LHC searches in both the $W^\pm+$MET and $W^+W^-+$MET channels.
\end{abstract}

\vspace{0.5cm}
\renewcommand{\thefootnote}{\arabic{footnote}}
\setcounter{footnote}{0}
\thispagestyle{empty}
\vfill
\newpage
\setcounter{page}{1}



\section{Introduction and PM Model Background}

While the exact nature of Dark Matter (DM) is presently unknown, models abound in the literature attempting to explain what it {\it might} be and how it could interact with the fields of the 
Standard Model (SM) to obtain it's relic density\cite{Planck:2018vyg}. One very attractive possibility is the vector boson/kinetic mixing (KM) portal scenario\cite{KM,vectorportal,Gherghetta:2019coi}, 
which posits (at a minimum) the existence of a new $U(1)_D$ gauge group, with a corresponding coupling $g_D$, whose gauge boson, the dark photon (DP),  
$V$\cite{Fabbrichesi:2020wbt,Graham:2021ggy,Barducci:2021egn,Alonso-Gonzalez:2025xqg,Jorge:2026bbs,Caputo:2026pdw}, directly couples to DM which carries a non-zero dark 
charge, $Q_D\neq 0$. 
The SM fields at tree level, on the other hand, do not couple to $V$ as they are all neutral under $U(1)_D$, \ie, they have $Q_D=0$. In the analysis presented below it will be assumed that the DP 
obtains a mass via the spontaneous breaking of this $U(1)_D$ via one or more dark Higgs fields which may or may not also transform non-trivially under the SM gauge symmetries\cite{Li:2024wqj}. 
Of course it is possible that particles carrying {\it both} the SM as well as $U(1)_D$ quantum numbers might also exist, which we have referred to as portal matter (PM) in earlier work\cite{Rizzo:2018vlb,Rueter:2019wdf,Kim:2019oyh,Rueter:2020qhf,Wojcik:2020wgm,Rizzo:2021lob,Rizzo:2022qan,Wojcik:2022rtk,Rizzo:2022jti,Rizzo:2022lpm,Bauer:2022nwt,Wojcik:2022woa,Carvunis:2022yur,Verma:2022nyd,Rizzo:2023qbj,Wojcik:2023ggt,Rizzo:2023kvy,Rizzo:2023djp,Rizzo:2024bhn,Ardu:2024bxg,Rizzo:2024kzu,Rizzo:2025tap,Tewary:2025vij,Rizzo:2025fsy}, these being either new scalars and/or new vector-like fermions (VLF) - this later choice being required in order to avoid unitarity, precision electroweak and Higgs width/production cross section  
constraints\cite{CarcamoHernandez:2023wzf,CMS:2024bni,Alves:2023ufm,Banerjee:2024zvg,Guedes:2021oqx,Adhikary:2024esf,Benbrik:2024fku,Albergaria:2024pji,Chen:2017hak,
Biekotter:2016kgi}. Depending upon their SM transformation properties, loops of these PM particles will then induce KM between $V$ and the SM $W_3$ and/or $B$ gauge fields via 1-loop vacuum 
polarization-like graphs which at low energies, far below the weak scale, will manifest themselves as the KM of the SM photon and $V$. In fact, in this class of models, it is more than likely that 
such states are {\it necessarily} a part of the physical spectrum at some scale in order to induce this desired KM. When $V$ and the DM are both light, roughly $\lsim$ a few GeV in mass, the 
strength of this induced KM, denoted by the parameter $\epsilon$, can be sufficiently large so as to allow the DM to achieve its relic abundance via the familiar thermal freeze-out processes, as in 
the case of WIMPS\cite{Arcadi:2017kky,Roszkowski:2017nbc,Arcadi:2024ukq}, while also simultaneously satisfying other experimental 
constraints.  This can happen because in such a setup, the charged fields of the SM will pick up a small coupling to $V$ via this KM with a strength $e\epsilon Q_{em}$, with $e$ being 
the usual charge of the proton.  In the traditional version of this model, no {\it other} additional interactions between the DP (and hence DM) with the SM are usually anticipated.

Recently, however, we have shown that in somewhat more UV-complete models of this general type, loops of PM fields in the form of VLF can in some cases generate new interactions for (at 
least some of) the SM fermions with the DP {\it beyond} that induced by KM in the form of, \eg,  `dark' dipole moment and/or dark charge form factors for these fields, even when they are 
electrically neutral. This can lead to some interesting phenomenological differences with respect to the more familiar KM setup\cite{Rizzo:2021lob,Rizzo:2023djp} 
while, \eg, still reproducing the DM relic abundance.  In earlier work, it was argued that the SM gauge bosons, $W^\pm,Z$, {\it cannot} pick up these types of couplings since the PM fields (which 
would come in mass degenerate $SU(2)_L$ representations) running in the loops which potentially generate them are all VL with respect to both $U(1)_D$ and 
the SM interactions so that Bose-symmetry plus charge conjugation invariance will force the sum of these loop diagrams to vanish. However, if {\it scalar} PM is instead realized, this 
no longer need be the case provided that the fields in any given loop are non-degenerate, as may easily happen for the $W$ where scalars having different third components of weak isospin 
in an $SU(2)_L$ representation are split in mass due to, \eg, their mixing with the SM Higgs doublet field. One reason for this is that these scalars can be relatively light so that the mixing induced 
effects on their masses can be sizable; in the fermion case, the mass limits and other experimental constraints are more restrictive, forcing the $SU(2)_L$ doublet members to be remain much more 
degenerate even after possibly mixing with the SM fields.

This possibility then directly leads to a new class of $WWV$ interactions which will be seen below to be of a similar qualitative nature to the 
long discussed (in the CP-conserving case we consider below) anomalous magnet dipole and electric quadruple moment triple gauge boson couplings of the $W$ (both in the SM and in numerous 
beyond the SM scenarios) when the DP is replaced by a photon or $Z$\cite{anom}. Of course, since the $Z$ boson couplings are generally diagonal, so that only a single field will propagate in a 
given loop at a time, as will be the case for the model considered here, this mechanism will not be effective in generating the corresponding $ZZV$-type couplings as the sum over loops will 
necessarily cancel via the above symmetry arguments.  It is to be noted that, unlike in the case of fermionic PM induced dark moment type couplings, no extension of the $U(1)_D$ dark gauge group 
is required in this setup although it would be certainly interesting to contemplate such more complex scenarios. 

In order to examine this possibility in some more detail, below we will resurrect a previously examined model of scalar PM\cite{Rizzo:2022jti} that was then motivated by the apparently large value of the 
$W$ boson mass (relative to the SM expectation) as measured by the CDFII collaboration\cite{CDF:2022hxs}, but now in a somewhat different region of the model parameter space. This 
setup augments the SM single Higgs doublet symmetry breaking sector by an additional {\it complex} isotriplet having $Y=0$, $\Sigma$, plus an additional complex singlet, $S$, both of which carry 
a non-zero value of the dark charge, \ie, $Q_D=1$, unlike the SM Higgs which has $Q_D=0$. Both of these new fields obtain vacuum expectation values, $v_{t,s} \lsim$ a few 
GeV, respectively, generating the DP mass as described above, but now with the expectation that $v_t\lsim v_s$ so that the tree-level contribution to the $\rho$ parameter, \ie, the 
Peskin-Takeuchi $T$ parameter\cite{Peskin:1991sw,Peskin:1990zt}, corresponding to the 
shift in the $W$ boson mass, is relatively suppressed in comparison to the expectations of our previous discussion. Thus these new scalars will not only act as PM but are also seen to be 
responsible as dark Higgs fields for the breaking of $U(1)_D$ at low mass scales. We note that the analysis of this rather simple setup is only meant to 
be suggestive of what might possibly occur in a much broader class of, somewhat unexplored, scalar PM scenarios that can lead to some distinctive phenomenology. We also note that other PM 
scalar scenarios, such as that introduced in Ref.\cite{Rueter:2020qhf}, can be shown to lead to qualitatively similar conclusions to those found below but are somewhat more complex to analyze 
due to their larger number of free parameters{\footnote{This results from the choice of there now being\cite{Rueter:2020qhf} three weak doublet $SU(2)_L$ Higgs scalar representations, two 
with  identical, non-zero values of $Q_D$.}}. We note, based on the above discussion, that we can always add extra VLF PM to this simple setup without there being any additional sizable 
contributions to these dark moment-like couplings of the $W$ to the DP.

What would be the best way to observe such dark moment $W$ couplings?  On-shell $W$ decay would be an ideal laboratory for these studies but the size of the data samples necessary to probe 
branching fractions in the range of interest to us ($\gsim$ a few $\cdot 10^9$) are unfortunately not likely to be available anytime in the foreseeable future. This is disappointing as ordinary KM and dark 
moment produced DP events can easily be shown to populate different regions of the three body phase space\cite{anom}. If the DP were instead just an ordinary SM $\gamma$ 
and/or $Z$, as in the case of the more familiar trilinear anomalous gauge 
couplings\cite{anom}, there are many other process which one could employ, with various combinations of the $W$, $\gamma$ and/or $Z$ being either on- or off-shell, \eg, $W^+W^-$ production 
at either a lepton or hadron collider or $W^\pm \gamma$ production at a hadron collider. All of these processes and the models for generating such couplings have been explored in the literature 
in great detail over the past decades and they are seen to generally (and obviously) rely on the relevant SM gauge fields having sizable couplings to the quarks and/or leptons in the initial state 
which initiate these reactions. In our case, however, processes such as $q\bar q, e^+e^- \to V^* \to W^+W^-$ are not very useful due to the strong coupling suppression arising from, \eg, KM, 
being already $O(\epsilon^2)\sim 10^{-7}$ at the {\it amplitude} level. However, a process such as $q\bar q' \to W^* \to WV$, followed by the hadronic decay of the $W$, while still having a suppressed 
amplitude, it is now only by a single power of $\epsilon$. In fact, in the familiar KM case where all charged fields have a coupling 
$e\epsilon Q_{em}$ to $V$ and in the limit that $m_V$ can be neglected, the differential cross sections for $q\bar q' \to W^*\to WV$ and $W\gamma$ production will be identical apart from an 
overall factor of $\epsilon^2$, arising from the same three graphs. In the case where KM is absent (or, more realistically, suppressed) and the only interaction of the $W$ and $V$ is through the 
$WWV$ loop-induced dark moment couplings we consider here, a $WV$ final 
state can still be produced {\it but} in now arises from only a single graph where the on-shell $V$ is emitted off the $W$ as $V$ no longer couples directly to any of the SM fermions.  Clearly the 
expectations for such a process will differ in detail from those of the familiar KM scenario although the final state in both cases will just be, \eg,  $W^\pm+$MET at the LHC assuming that the DP decays 
to DM. Does this process occur at a significant rate which is large enough to be observable at the HL-LHC and how do the two distinct models compare? Does the direct production of PM scalars in the 
loop at the LHC provide better access to this physics and do they produce more significant signatures that can be probed at present? These amongst others are some of the related questions we 
hope to address in the discussion below. 

The outline of this paper is as follows:  After this Introduction, in Section 2 we will present/review the specific details of the model discussed above wherein the scalar PM fields consist of a complex, 
$Y=0$, $SU(2)_L$ triplet plus complex singlet combination, both with $Q_D=1$, which will also act as dark Higgs generating the DP mass. The present value of the $W$ mass and the needs of 
the current study are seen to push us into a previously unexamined parameter space regime for this setup. Specific expressions for the relevant scalar masses and mixings as will be important for 
the calculations of the dark moments of the $W$ will be presented to leading order in the small vev ratios. In Section 3, we will discuss the various dark moment form factors entering the effective 
$W^*WV$ vertex, write expressions for them in the model under discussion, and then provide the resulting cross section for the $q\bar q'\to W^*\to WV$ process. In Section 4, the 
event rate for this process at the LHC will be analyzed and compared to that obtained in the more convention KM scenario. Although found to be larger by more than an order of magnitude than that 
in the corresponding KM case, the dark moment induced cross section is still very likely too small to be observed, even at the HL- LHC, due to the much larger SM backgrounds arising from, \eg, 
$WZ$ production followed by $Z\to \bar \nu \nu$. In fact, the additional new physics backgrounds to this process induced by the direct production of the PM scalars in this model are shown to 
be potentially large and so we then study them in more detail instead within this new parameter space regime. For these PM scalar production processes, for the simplest final states, such as 
$W+$missing transverse energy (MET) and the $W^+W^-+$MET, search constraints already exist from the LHC and are compared with the scalar PM model predictions. Although much more serious 
simulation-level studies are certainly needed, these results indicate that the HL-LHC will likely be able to probe this direct production possibility over a good fraction of the model parameter space 
for interest. Finally, a discussion of the results and our conclusions are presented in Section 5.


\section{Review of Scalar PM Model Basics}

As noted in the Introduction, the basic model assumptions that we will make below are as follows: ($a$) The SM gauge group is augmented by an additional 
$U(1)_D$ factor corresponding to the $\lsim 1$ GeV, dark photon, $V$. The SM fields all have $Q_D=0$, so that before any loop effects/KM and/or mass mixing they do not couple to $V$.  
($b$) The Higgs sector, in addition to the usual SM doublet with $Q_D=0$, \ie, $\Phi$,  also consists of a complex singlet, $S$, with $Q_D=1$, having vev, $v_s$, as well as a complex $Y=0$, isotriplet, 
$\Sigma$, which has the same dark charge, $Q_D=1$, and whose neutral member also obtains a vev, $v_t$. As noted above and as is very well-known this leads to a positive tree-level shift 
in the $\rho$ parameter, $\delta \rho=4v_t^2/v_d^2$, where $v_d\simeq 246$ GeV is the familiar vev of the SM doublet $\Phi${\footnote {Note that if $v_t<1$ GeV, which will certainly be the case 
here, then $\delta \rho <10^{-4}$.}}. The fact that $\Sigma$ is complex implies that its oppositely 
charged $T_3=\pm 1$ components, $\Sigma_{1,2}$, are {\it not related} to one another, having distinct masses, and that the $T_3=0$ component, $\Sigma^0$, is also necessarily 
a complex field. While the triplet vev, $v_t$, and the singlet vev, $v_s$, which we'd naively expect to be of comparable size $\lsim 1$ GeV, will {\it both} contribute to the mass of the DP, below we 
will consider the parameter space regime where $v_t <v_s$ since we want the $W$ (at least at the tree level) to obtain the vast bulk of its mass from the SM vev, $v_d$, to a rather high 
precision given comparisons to and agreement of the current experimental measurements with SM expectations\cite{Bozzi:2024lox}.  ($c$) While other PM which are VLF in degenerate $SU(2)_L$ 
representations may also be present, their contributions to the dark moment-type couplings of the $W$ as considered here will sum to zero as discussed above and so would not play very much of a 
direct role in the discussion below. 

Closely following our previously analysis\cite{Rizzo:2022jti}, we can decompose the real and imaginary parts of these three {\it complex} Higgs fields together with their vevs in the following manner: 
\begin{equation}
\Sigma = \begin{pmatrix} \Sigma_1^+ \\\ \Sigma^0=\frac{\sigma+v_t+i a_t}{\sqrt{2}} \\\ \Sigma_2^-\end{pmatrix}, ~~~
S = \begin{pmatrix} \frac{s+v_s+i a_s}{\sqrt{2}} \end{pmatrix}, ~~~
\Phi = \begin{pmatrix} \phi^+ \\\ \phi^0=\frac{h+v_d+ia}{\sqrt{2}} \end{pmatrix} 
\end{equation}
with $v_{t,s,d}$ being the relevant vevs of the triplet, singlet and SM doublet Higgs fields, respectively, as noted above and where we have explicitly shown the real and imaginary components of the 
fields with $Q_{em}=0$. The interactions of these scalar fields with themselves and with each other are given by the most general scalar potential, $U$, that is allowed by both the 
SM and $U(1)_D$ gauge symmetries, \ie, 
\begin{equation} 
\begin{aligned}
U = & -m^2 \Phi^\dagger \Phi -m_\Sigma^2 Tr(\Sigma^\dagger \Sigma) -m_S^2 S^\dagger S +\lambda ( \Phi^\dagger \Phi)^2 +\lambda_S( S^\dagger S)^2  +\lambda_1 [Tr( \Sigma^\dagger \Sigma )]^2 + \lambda_2 Tr [(\Sigma^\dagger \Sigma )^2] \\& +\kappa \Phi^\dagger \Phi S^\dagger S + [\lambda_{4\Phi} \Phi^\dagger \Phi +\lambda_{4S} S^\dagger S]~Tr(\Sigma^\dagger \Sigma)+\lambda_5\Phi^\dagger \Sigma\Sigma^\dagger \Phi +\tilde \lambda \Phi^\dagger (\Sigma S^\dagger +\Sigma^\dagger S)\Phi\,.
\end{aligned}
\end{equation}

A few comments are now in order: ($i$) While all of the quartic couplings as well as the magnitudes of the various vevs play important phenomenological roles, we specifically note that the 
presence of the $\tilde \lambda$ term which is only possible since {\it both} $S$ and $\Sigma$ have the same dark charge, $Q_D=1$.  ($ii$) If CP remains a good symmetry (which we will assume 
here) then the mass eigenstates of the CP-even and CP-odd neutral fields can be separately determined after minimization of the potential; Ref.\cite{Rizzo:2022jti} describes this procedure in 
detail. ($iii$) We will denote the ratios of the small vevs to that of the SM vev by $x_{t,s}=v_{t,s}/v_d\sim O(10^{-3}-10^{-2})<<1$, so that the tree-level value of the $\rho$ parameter is just given by 
$\rho=1+4x_t^2\simeq 1$ to a very high precision; we also will define the ratio of the two small vevs themselves, $t=v_t/v_s < 1$. In most of our discussions it will generally be sufficient for our purposes 
to work to leading order in the two small parameters, $x_{t,s}$.  ($iv$) Before any mass and/or kinetic mixing, the DP's mass is given by $m^2_V=g_D^2(v_s^2+v_t^2)=g_D^2v_s^2(1+t^2)$, 
with $g_D$ being the $U(1)_D$ dark gauge coupling as defined above. ($v$) It is to be noted that both of the new neutral scalar fields that obtain vevs have $T_3=0$, \ie, their third components 
of weak isospin vanish, so that their mixing does not induce any off-diagonal neutral current couplings to the SM $Z$ or, in fact, to the $V$ since they also both have the same value of $Q_D=1$.

With these comments in mind, it was found that to lowest order in the parameters $x_{t,s}$, of the neutral CP-odd fields, one of them, $a \simeq G_Z$, becomes the Goldstone boson for the SM 
$Z$ while one linear combination of the remaining CP-odd fields, $G_V\simeq a_s c_\phi+a_t s_\phi~$ with $t=t_\phi=s_\phi/c_\phi=v_t/v_s$, becomes the Goldstone boson for the the DP field, 
$V$. Here, we have defined, in familiar notation, $s_\phi(c_\phi)=\sin \phi(\cos \phi)$, as the mixing angle between the weak eigenstate and mass basis. The remaining CP-odd mass eigenstate is 
then just the orthogonal state $A\simeq a_t c_\phi-a_s s_\phi~$ and is found to be relatively heavy, \ie,  
\begin{equation}
M_A^2=\frac{\tilde \lambda v_d^2}{2\sqrt 2} \left( t+\frac{1}{t} \right)\,,
\end{equation}
to lowest order in the $x_{t,s}$ expansion.  Note that we will sometimes employ the $G_V$ notation to represent the longitudinal mode of the DP, $V_L$, in the Goldstone Boson Equivalence 
Theorem\cite{GBET} limit in the discussion below.  Similarly, considering the neutral CP-even states, to the same order in $x_{t,s}$ parameters, one can be identified with the $h_{SM}\simeq h$ 
state at $\simeq 125$ GeV observed at the LHC, which has very little mixing with the other states to this order, while a second, 
$h_D\simeq s c_\phi +\sigma s_\phi$ with a mass near $\sim1$ GeV, essentially set by $v_s$, 
is identified with the dark Higgs. The orthogonal state, $H \simeq \sigma c_\phi -s s_\phi$, is found to be heavy on the other hand; explicitly one obtains the following expressions to lowest order 
in the small parameters for these three masses: 
\begin{equation} 
\begin{aligned}
m_{h_{SM}}^2 \simeq & ~2\lambda v_d^2\\
m_H^2 \simeq &~m^2_A =m_{\cal H}^2=\frac{\tilde \lambda v_d^2}{2\sqrt 2}~\left(t+\frac{1}{t} \right )\\
m_{h_D}^2 \simeq & ~\frac{2\lambda_S v_s^2}{1+t^2}~\left(1-\frac{\kappa^2}{4\lambda \lambda_S} \right)\,,
\end{aligned}
\end{equation}
where we see that the SM Higgs mass is effectively unaltered to this order. The dark Higgs mass, $m_{h_D}$, is seen to be similar to the case of the simple singlet dark Higgs model except for 
the presence of the additional overall $(1+t^2)^{-1}$ factor.  Recall that it is necessary to suppress $h_{SM}-h_D$ mixing to satisfy the branching fraction constraint on the invisible decay of the 
SM Higgs, \ie, $B_{inv}\leq 0.11$\cite{Invisible}, which here will arise from the decay modes $h_{SM}\to 2h_D,2G_V$, assuming that both the dark Higgs and the DP decay invisibly or 
to unreconstructed final states. It is important for later discussion that we observe that the heavy CP-even and CP-odd states, \ie, $H$ and $A$, are found to be essentially 
degenerate to this order, having identical admixtures of the singlet scalar;  thus, to this order in the small parameters, we can go back and treat just the combination ${\cal H}=(H+iA)/\sqrt 2$ as a 
single complex field of mass $m_{\cal H}$ when performing the calculations below. Similarly, at $\sim $ TeV scales, we can think of $h_D$ as being almost massless and combine it with $G_V$ 
to form a {\it complex} state which we'll still refer to as $h_D$ for convenience. We note in passing that since both the neutral complex fields, $h_D, {\cal H}$, have $T_3=0$, they will not couple to 
the SM $Z$ boson which reduces the number of channels that can be used to access these particles directly at a collider as will be important in the discussions later below.

Finally, the diagonalization of the mass-squared matrix for the charged scalar fields is seen to yield one null eigenvalue corresponding to $\phi^+ \simeq G_W$, the SM $W$'s Goldstone boson, 
and in addition, to leading order in the small parameters, the masses of both $\Sigma_{1,2}$:
\begin{equation} 
m_{1,2}^2 =\left( \frac{\tilde \lambda}{2\sqrt 2} \frac{1}{t} \mp \frac{\lambda_5}{4}\right)v_d^2\,,
\end{equation}
where $\lambda_5>0$ implies the ordering $m_1<m_2$. We also observe that we must require that $\tilde \lambda >\lambda_5 t/\sqrt 2$ in order to avoid turning $\Sigma_1$ into a tachyonic 
state.  As was previously noted, we see that the squares of the masses of the three complex scalars, $m_{1,2}^2$ and $m_{\cal H}^2$, are, to lowest order in $x_{t,s}$, simply related by 
\begin{equation} 
\frac{m_{\cal H}^2}{m_1^2+m_2^2}= \frac{1+t^2}{2}\,,
\end{equation}
which for $t \lsim 1$, as we will assume here, is expected to be roughly $\lsim O(1)$. Considering the results above we then might expect that $m_{h_D}<<m_1<m_{\cal H}<m_2$ which will be of 
some significance when we perform the loop calculations below; note, however, that either mass ordering between ${\cal H}$ and $\Sigma_2$ is in general possible in the $t\lsim 1$ region.

Note that once we assume that the masses of $h_D$ and $V$ are $\lsim 1$ - a few GeV, then the most important model parameters become just the three scalar masses $m_{1,2,{\cal H}}$ from 
which the value of $t$, and thus the mixing angle $\phi$ (and so all of the various gauge interactions) can be then be determined to lowest order in the ratios $x_{t,s}$. However, as was also 
previously noted\cite{Rizzo:2022jti}, some of the the pure scalar sector interactions, controlled by the multiple quartic couplings, $\lambda_i$, appearing in the potential above will still remain 
somewhat unconstrained and so naturally can likely be $O(1)$ thus also be important when considering interactions between the scalars themselves and the corresponding collider production 
processes that can result from the trilinear couplings proportional to the largest vev, $v_d$.

\section{Generating $W$ Dark Moments and LHC Signatures}

\begin{figure}[htbp] 
\hspace*{0.3cm}\centerline{\includegraphics[width=5.5in,angle=0]{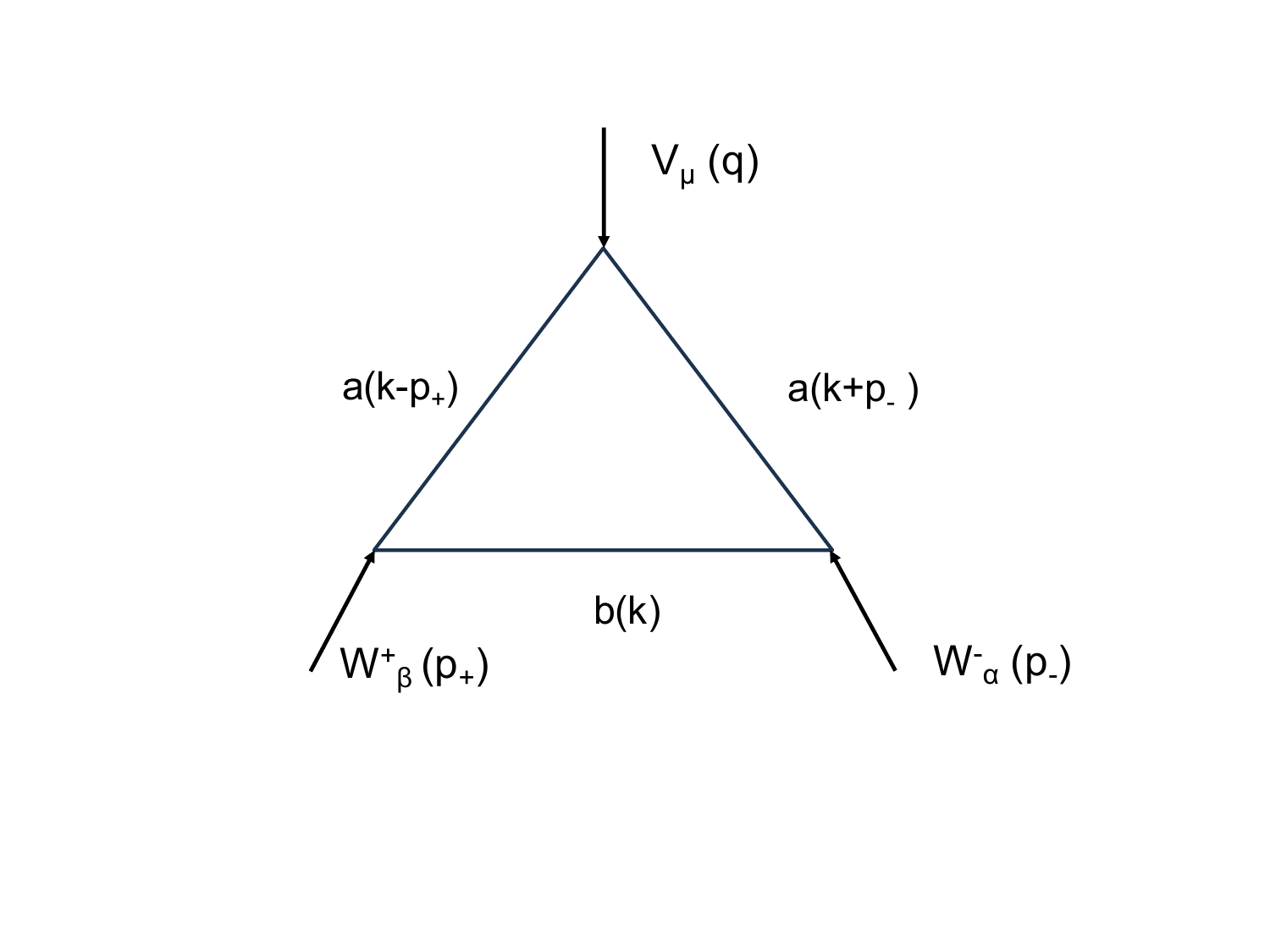}}
\vspace*{-2.0cm}
\caption{Generic triangle graph of PM scalars, labeled by the pairs $(a,b)$ discussed in the text, for our PM model that generate the dark moment-like couplings of the SM $W$ to the dark photon, $V$. 
The momentum in the loop as labeled is seen to flow counterclockwise in this Figure. }
\label{fig-t}
\end{figure}

Unlike in the familiar KM scenario wherein all the SM charged fields pick up an effective tree-level coupling to the DP, $e\epsilon Q_{em}$, here, in the case where the KM is suppressed or absent, 
only the 
$W$ can potentially pick up a sizable loop-induced coupling to $V$ since any SM fermionic couplings to these scalars which occur via mixing are necessarily quite suppressed due to the bound arising 
from the invisible width of the Higgs from the LHC\cite{Invisible}. Given the set of scalars and their masses described above it is easy to contemplate how these dark $WWV$ couplings may 
be generated - qualitatively in a somewhat similar manner to how the fields of the Two Higgs Doublet Model can lead to contributions to the more familiar anomalous $WW\gamma$ couplings. An 
important difference here is to remember that even the electrically neutral PM scalars of our model have a dark charge so that they can also couple directly to the DP. While a $W^\pm$ must couple 
to either of $\Sigma_{1,2}^\pm$, it must also couple to ${\cal H}$ or $h_D$ in such a manner so as to conserve the dark charge $Q_D${\footnote {We remind the Reader that, as noted previously, 
in what follows we will for simplicity employ $h_D$ to denote the {\it complex} light state, roughly in the mass range $\sim 0.1-1$ GeV.}}.  Here the relevant family of triangle graphs with 
both (electrically) charged and neutral scalars, generically labeled as $(a,b)$, one being electrically charged with the second neutral, running clockwise in these loops, is shown in Fig.\ref{fig-t}. 
A corresponding crossed graph will also exist wherein the two $W$'s are interchanged and which then introduces an overall minus sign in the $WWV$ amplitude\cite{anom} and, due to electric 
charge conservation, necessarily interchanges the roles of the states, 
$a \leftrightarrow b$, so that if $a$ and $b$ were to be degenerate (or simply to be the same state as in the case of potential $ZZV$ couplings) the sum of these two graphs would vanish. This explains 
why the scalar-induced CP-conserving dark moment-like couplings for the SM $Z$ will vanish as was discussed above since no off-diagonal $Z$ couplings are allowed at tree-level. It also tells us 
to expect that the size of these dark moment couplings will be very roughly set by the mass-squared splittings among these various scalars (as well as their mixings) roughly scaled to that of the 
square of the $W$ boson mass.  Of course this weak isospin breaking contribution to these mass-squared differences are due to the same vev that generates the $W$ mass, \ie, $v_d$.

Note that given the direction of the momentum flow and the charge labels for the $W$'s in Fig.~\ref{fig-t}, we see that there are only 4 choices for the charged plus neutral scalar pairs that 
can appear in the loop, \ie,   $(a,b)=(\Sigma_2^-,{\cal H}/h_D)$ and $(a,b)=({\cal H}/h_D,\Sigma_1^+)$ with the weighting between the ${\cal H}$ and $h_D$ contributions being determined by 
the values of the mixing factors,  $c_\phi^2$ and $s_\phi^2$, discussed above. The crossed graphs then correspond to the same pairings but with the roles of $a$ and $b$ interchanged as well as 
the relative additional overall change of sign. Note that with this interaction structure these dark moment couplings for the $W$ will all be finite quantities as is clear from the straightforward 
application of dimensional regularization as any remaining poles will vanish between the original and crossed graphs.

In earlier work\cite{Rizzo:2022jti}, a scan was performed over the parameter space of this model subject to several additional constraints, in particular, the requirement that the calculated value of 
$\epsilon$ was sufficiently large within the pure KM framework to recover the DM relic density. If we drop this constraint (and also allow for the new requirement $t<1$), as we assume that the 
KM contributions to the process at hand are absent, or at the very least, sub-dominant, a somewhat different but slightly larger parameter space becomes open to us. This is of course subject to 
the remaining constraints, such as those arising from both unitarity of the 
$\lambda_i$'s and the direct search constraints on the new scalar PM states to be discussed below. In particular, overall, somewhat larger scalars masses aree allowed; however, when identifying 
benchmark parameter points in this space for further study we will remain rather conservative. We further note that varying the value of $m_{h_D}$ anywhere within the 0.1-1 GeV range only leads 
to alterations in our numerical results below at the sub-percent level and so this variation can be neglected and we set $m_{h_D}=1$ GeV in the relevant loops for definiteness. Below, $m_0$ will 
label either the $h_D$ or ${\cal H}$ mass depending upon which one appears in the loop diagram.  Table 1 gives the parameters for the two sample benchmark sample we will employ in the 
analysis below for demonstration purposes. However, at the semi-quantitative level of roughly a factor of 2 or so, one finds that much of the $t<1$ model space gives quite similar results to what we 
will discuss below. Note that we will also make use of generalized versions of these two benchmark models that will be further analyzed below as part of our collider studies where $m_1$ is 
treated as a free parameter but all of the scalar mass ratios (and hence $t$) are held fixed.

\vspace{0.5cm}
\begin{table}
\caption{Benchmark Model Points}\label{qtab}
Benchmark model point properties: PM scalar masses in GeV for those states appearing in pairs in the triangle loop graph Fig.~\ref{fig-t} and generating the dark moment couplings as described 
in the text. In the generalized versions of these two benchmarks also discussed later in the text, the values of $m_1$ are treated as free parameters while that of $t$ and all of the scalar mass 
{\it ratios} are held fixed.
\begin{center}
\begin{tabular}{ l c c c c}
\hline
\vspace{-0.1cm}\\
BM & $m_1$ &  $m_2$ & $m_{\cal H}$  & $t$ \vspace{0.1cm}\\
\hline 
\vspace{-0.05cm}\\
BM1 & 330 & 520 & 470 & 0.406\vspace{.01cm}\\
BM2 & 350 & 600 & 580 & 0.628\vspace{.01cm}\\
\hline 
\vspace{0.5cm}\\
\end{tabular}
\end{center}
\end{table} 

The general tensor structure for the CP-conserving trilinear $WW\gamma/Z$ vertex allowing for all external particles to be off-shell has been given by Eq.(3) in the second paper of Ref.~\cite{argyres}; 
we will make use of a modified version of this expression in the analysis that follows. Here we must remember several things in order to employ this result for the $WWV$ vertex, specifically for 
the process we consider: First, unlike for the case of the anomalous SM-like $WW\gamma/Z$ coupling, there are {\it no}  tree-level contributions to this vertex in the absence of KM. Second, since 
the final state $W,V$ are both on-shell, the ratio $m_V^2/m_W^2 <<1$ as well as the initial state quark masses can be neglected, so we can drop all terms in this general coupling structure that 
are proportional to the vectors $q^\mu,p_+^\alpha$, or $p_-^\beta$ as well as those proportional to $q^2=m_V^2$. The reasons for this are clear: both $q^\mu$ and $p_-^\beta$ are to be 
contracted into the corresponding polarization vectors for external gauge fields and so these terms vanish identically. Terms proportional to $p_+^\alpha$ will be contracted with the initial quark current 
which is conserved in the massless limit as is considered here so these terms will also vanish. Further, we can safely numerically neglect terms which are suppressed by factors of $m_V^2/m_W^2$ 
so that all terms proportional to $q^2$ can also be dropped. 

With these results in hand, employing momentum conservation and with the external momentum flow now altered to correspond to a single incoming, off-shell $W$ and an {\it outgoing} $W,V$ pair 
in the final state with both the $W$ and $V$ now on-shell, as per the $q\bar q' \to W^* \to WV$ process under investigation, the relevant effective $WWV$ vertex for the case at hand can be written as
\begin{equation}
\Gamma_{\alpha \beta \mu}=g_D\Big[\big(a_0 g_{\alpha \beta}+a_1 q_\alpha q_\beta\big)(p_+ +p_-)_\mu-a_2 g_{\alpha \mu} q_\beta+a_3g_{\beta\mu}q_\alpha\Big]\,,
\end{equation}
where in our setup all the form factors will vanish, $a_i=0$, at tree-level, being purely loop-induced. Note that while $a_{0,2,3}$ are dimension-4 quantities, $a_1$, on the other hand, is instead seen to 
be dimension-6. Given the coupling structure from the second paper of Ref.~\cite{argyres} as noted above it is clear that these form factors are not completely independent quantities thanks to 
gauge invariance. For example, some simple algebra shows us that for the case at hand, \eg, $a_3(\hat s)=a_2(\hat s)-(\hat s-m_W^2)a_1(\hat s)$, where $\hat s$ is the squared 
partonic center of mass energy as is relevant for colliders. This is the only off-shell momentum-squared for the process we consider since the final state $W$ and $V$ are both produced on-shell, \ie, 
only the virtual intermediate $W$ will be off-shell here. We have verified numerically that this relationship indeed holds true in the numerical study below. From this same Reference, we see that if 
we write their $\kappa=1+\Delta \kappa$ then we find that in this `anomalous coupling' notation we can relate $a_1=-\lambda/m_W^2$ and $a_2=\Delta \kappa+\lambda$, so making contact with other 
existing literature\cite{anom} on the more familiar $WW\gamma/Z$ anomalous triple gauge couplings. 

In terms of these $a_i$, the LO subprocess $q\bar q' \to W^*\to WV$ differential cross section is found to be given by
\begin{equation} 
\frac{d^\sigma_{WV}}{dp_T dy}=\frac{g_D^2}{g^2}~\Big(\frac{G_F^2m_W^4}{3\pi \hat s^2}\Big)~\Big[(\hat s-m_W^2)^2+m_W^2\Gamma_W^2\Big]^{-1}\cdot {\cal J}\cdot {\cal F}^2~\cdot {\cal C}\,,
\end{equation}
where $y$ is the usual rapidity, $p_T$ the transverse momentum of the $W$ or $V$, $G_F$ is the Fermi constant, $\Gamma_W$ is the $W$ total width, 
$\simeq 2.09$ GeV\cite{ParticleDataGroup:2024cfk}, ${\cal F}^2$ is a numerical factor arising from the loop graphs to be discussed below and ${\cal J}$ is the appropriate familiar Jacobian 
factor\cite{Eichten:1984eu} as needed for these final state variables. Note that in the numerical analysis below we will assume the overall coupling ratio value of $g_D/g=1$. Further, we see that 
${\cal C}$ is the sum of several terms, 
${\cal C}=\Sigma_{j=1,9}~C_j$, that can be most easily expressed in terms of the usual subprocess Mandelstam invariants, $\hat s,\hat t,\hat u$ (that are subject to the usual constraint that 
$\hat s+\hat t+\hat u=m_W^2$) and with the various $a_i$ being just the form factors from the coupling structure defined above, \ie, 
\begin{eqnarray}
C_1&=&-\hat u \hat t ~\Big(\frac{a_1}{2m_W}\Big)^2 ~(\hat s+m_W^2)(\hat s-m_w^2)^2  +\hat u \hat t~\frac{a_3}{2}~(3a_3-2a_2)\nonumber\\
C_2&=& \frac{a_2a_3}{4}~\Big(\frac{\hat s}{m_W^2}-1\Big)~\Big( \hat t(\hat t-m_W^2)+\hat u(\hat u-m_W^2)\Big)+a_2(a_2-a_3)~\frac{\hat s}{4m_W^2}~(\hat s-m_W^2)^2      \nonumber\\
C_3&=& \frac{-a_1a_2}{4m_W^2}~\Big(\hat t(\hat s-\hat t+m_W^2)+\hat u(\hat s-\hat u+m_W^2)+\hat s(\hat s-m_W^2)\Big)~(\hat s-m_W^2)^2\nonumber\\
C_4&=& a_1a_3~\hat u \hat t~\Bigg[\hat s-m_W^2-\frac{1}{4}\Big(\frac{\hat s}{m_W^2}-1\Big)~(\hat s+3m_W^2)\Bigg]                   \nonumber\\
C_5&=& \frac{a_0^2}{2m_W^2}~(\hat s-m_W^2)~\Big(2m_W^2\hat s-(\hat t-m_W^2)(\hat u-m_W^2)\Big)        \nonumber\\
C_6&=& \frac{a_0a_1}{2}~(\hat s+m_W^2)~\Bigg[4\hat u \hat t-\Big( \frac{\hat s}{m_W^2}-1\Big)~\Big(\hat t(\hat t -m_W^2)+\hat u(\hat u-m_W^2)-\hat s(\hat s-m_W)\big)\Bigg]\nonumber\\
C_7&=& -\frac{a_0}{2}(a_3-a_2)\Big(\hat t(\hat s-\hat t+m_W^2)+\hat u(\hat s-\hat u+M_W^2)+2\hat s(\hat s-m_W^2)\Big)\nonumber\\
C_8&=&-\frac{a_0a_3}{4m_W^2}~(\hat s+3M_W^2)~\Big(\hat t(\hat t-m_W^2)+\hat u(\hat u-m_W^2)-\hat s(\hat s-m_W^2)\Big)\nonumber\\
C_9&=-&\frac{a_0a_2}{8m_W^2}~(\hat s-m_W^2)~\Big((\hat u-m_W^2)(\hat s -\hat t+m_W^2)+(\hat t-m_W^2)(\hat s-\hat u+m_W^2)+\hat s(\hat s+3m_W^2)\Big)\,.
\end{eqnarray} 

In order to proceed further, the expressions for the $a_i$ themselves are required; here we will employ for simplicity the dimensionless quantities 
$r_{a,b}=m_{a,b}^2/m_W^2$, where $(a,b)$ label the scalars in the loop as defined above,  $r_p=\hat s/m_W^2$ and define ${\cal F}=g^2/4\pi^2=\sqrt 2 G_Fm_W^2/\pi^2 \simeq 1.08\cdot 10^{-2}$ 
as a common overall constant arising from the $W$ couplings and various loop factors that has been extracted form these expressions and moved into the cross section as defined above. Explicitly, 
we find that we can express the $a_i$ as a set of two-dimensional parameter integrals, \eg, 
\begin{equation}
m_W^2a_1{\cal F}^{-1}=\int_0^1 dw~\int_0^1 du ~\frac{uw^3(1-w)1-u)}{r_aw+(1-w)\big(r_b-uwr_p-w(1-u)\big)}\,,
\end{equation}
where here $u,w$ are just the usual Feynman parameters. Similarly, we have that 
\begin{equation}
a_{(2,3,0)}{\cal F}^{-1} =\int_0^1 dw~\int_0^1 du ~\Big(-uw^2,  -(1-u)w^2  ,\frac{-w}{2}(1-w)\Big) ~ {\rm log}\Big[r_aw+(1-w)\big(r_b-uwr_p-w(1-u)\big)\Big]\,,
\end{equation}
all of which we can evaluate numerically for the different choices of $(a,b)$. We can then combine these results together with the appropriate mixing angle weightings for ${\cal H},h_D$ employing 
the two BM models, thus obtaining their final values and then feed these into the cross section expressions above. Numerically, we find that the contribution from $a_0$ is relatively quite suppressed 
in comparison to those from the other two dim-4 form factors, $a_{2,3}$, and consequently makes essentially no contribution to the cross section for the two BM examples at hand. The dim-6 form 
factor $a_1$ is also generally suppressed but, unlike in the case of the the dim-4 $a_0$, always appears in conjunction with additional powers of, \eg, $\sim \hat s/m_W^2$ which essentially 
compensates for this relative suppression.

\section{Dark $W$ Moments at LHC: Cross Section and PM Direct Production}

\begin{figure}[htbp] 
\centerline{\includegraphics[width=5.0in,angle=0]{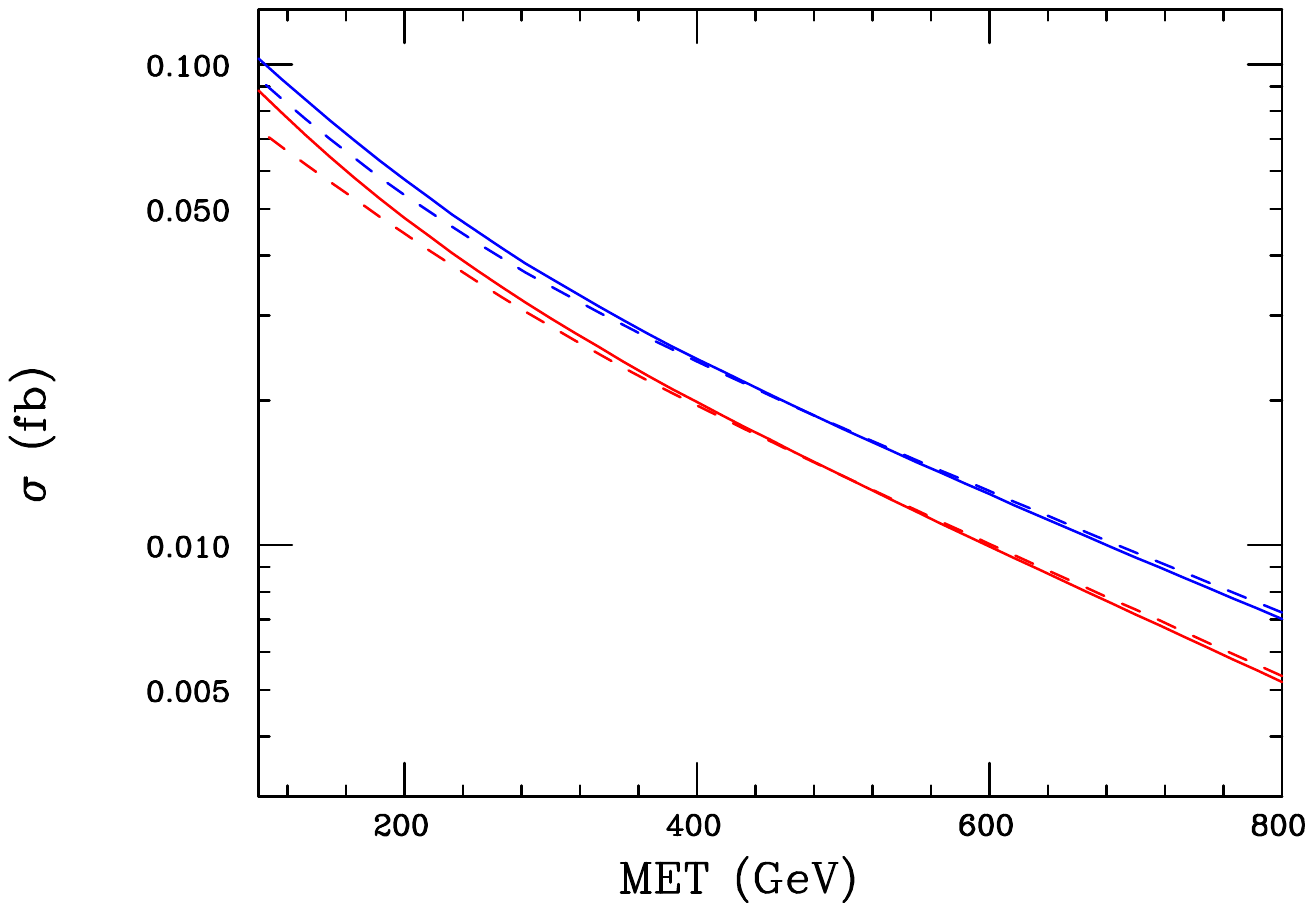}}
\vspace*{-1.3cm}
\caption{Production cross section for the $W^\pm V$ final state in fb at the $\sqrt s=13$ TeV (lower red) and 14 TeV (upper blue) LHC as a function of the cut on the missing transverse energy (MET)  
associated with the invisible dark photon. The solid (dashed) curves correspond to the benchmark model BM1 (BM2) as discussed in the text.}
\label{fig0}
\end{figure}

We now have all the pieces necessary to determine the dark moment induced $WV$ production rate;  Fig.~\ref{fig0} shows this cross section for $W^\pm V$ at the 13 and 14 GeV LHC 
as a function of the cut on the missing transverse energy (MET) for both benchmarks BM1 and BM2 employing the numerical analysis above. Here we see these benchmarks yield qualitatively 
quite similar results although the shapes are slightly different in detail due to the modest differences in the form factors and their relative weightings in the final cross section. Other points in this 
parameter space will also yield results similar to these two benchmarks even for somewhat larger scalar masses. For example, scaling all the masses of BM2 upward by a factor of 1.5 (thus 
leaving the value of $t$ unaltered) changes the resulting cross section for BM2 by roughly a factor of $\lsim 20-30\%$.

The first question one might ask is how these cross sections compare to the more conventional `pure' KM prediction for the analogous $q\bar q' \to W^*\to WV$ process. 
In order to make this comparison easily, we can employ Ref.\cite{Kidonakis:2025jve} who have recently obtained the corresponding $W^\pm\photon$ production cross section at 
mixed NNLO/N$^3$LO order including soft gluon corrections. Summing over the contributions of both $W^\pm$ final states for the case of the $\sqrt s=13$ TeV LHC with a $p_T$ cut on the photon of 
25 GeV, they find the central value of (ignoring the estimated errors) $\sigma_\gamma \simeq 110$ pb. If we again neglect its mass, the corresponding DP production cross section is obtained by 
simply scaling 
by $\epsilon^2 \simeq 10^{-7}$, or so, yielding a cross section roughly $\sigma_V \simeq 11~(\epsilon^2/10^{-7})$ ab. On the other hand, BM1 yields a cross section of roughly $\simeq 190$ ab at 
LO (scaled by a $K$-factor of 1.3) with this same (now MET) cut which is seen to be more than an order of magnitude larger while BM2 yields a cross section of a quite similar size as can be seen from 
Fig.~\ref{fig0}. Unfortunately, in either 
case, not only is this signal quite swamped by, \eg, the SM $W^\pm Z, Z\to \nu \bar \nu$ background but also, in most parameter space regimes as we'll see below, by that which arises from the 
{\it direct} production of the PM scalars that can run in the loops. In the case of fermionic dark moments, as seen earlier\cite{Rizzo:2021lob}, this additional background would {\it not} occur 
as the corresponding VL 
PM fermions are much more massive than the typical PM scalars by at least a factor $\simeq 3-5$ and so will have relatively suppressed pair production cross sections. Furthermore, 
this large SM background is itself seen to be responsible for the relatively modest LHC Run 2 constraints\cite{ATLAS:2024rlu} on any excess in the $W/Z \to ~hadrons$ + MET 
channel arising from new physics sources. From the model-independent results presented in Fig. 6 of this ATLAS analysis (to which we will return below) we see that, even employing the 
most promising class 
of signal events (those with high purity merged jets) and also ignoring possible acceptance and efficiency differences, the upper bound on the cross section for this process is roughly 
$\simeq 7$ fb assuming a cut MET$>250$ GeV.  This is seen to be more than two orders of magnitude larger than our expected DP signal rate as is seen from Fig.~\ref{fig0}. Clearly, assuming no 
very significant improvements in the analysis, the $WV$ final state induced by the dark moment couplings of the $W$ will very likely remain inaccessible even at the HL-LHC. 

Now, of course, the rate for $WV$ production will be substantially larger at the FCC-hh due to both the much higher collision energies and larger integrated luminosities available and so will very 
likely be accessible. However, at that point, all of the PM scalars that have been up to now indirectly responsible for this process via loops can be produced on-shell and directly studied. In fact, 
some of them, \eg, $\Sigma_1^\pm$, being relatively light in much of the model parameter space, may already be directly accessible at the LHC. This possibility was briefly discussed in our earlier 
work\cite{Rizzo:2022jti} although now their 
decay signatures could differ somewhat in the present region of parameter space, as represented by the benchmark models, from those previously assumed as we are now living in 
the $t<1$ regime. Can the HL-LHC directly probe the underlying mechanism that generates these dark moment interactions instead of the interactions themselves? If the PM had been a VLF, present 
LHC search reaches may already make this an unlikely possibility. However, in the present case, we are instead dealing with lighter PM scalars which may be more promising. To this end, we can ask 
how large the PM cross sections themselves are compared to the dark moment results as well as the SM backgrounds. Clearly this will depend to some extent upon our specific location in the 
model parameter space. In all of these considerations it is important to recall that at leading order in $x_{t,s}$ none of the PM scalars will directly couple to the SM fermions or to gluons, which are 
the likely initial states in any collider production process.

\begin{figure}[htbp] 
\centerline{\includegraphics[width=5.0in,angle=0]{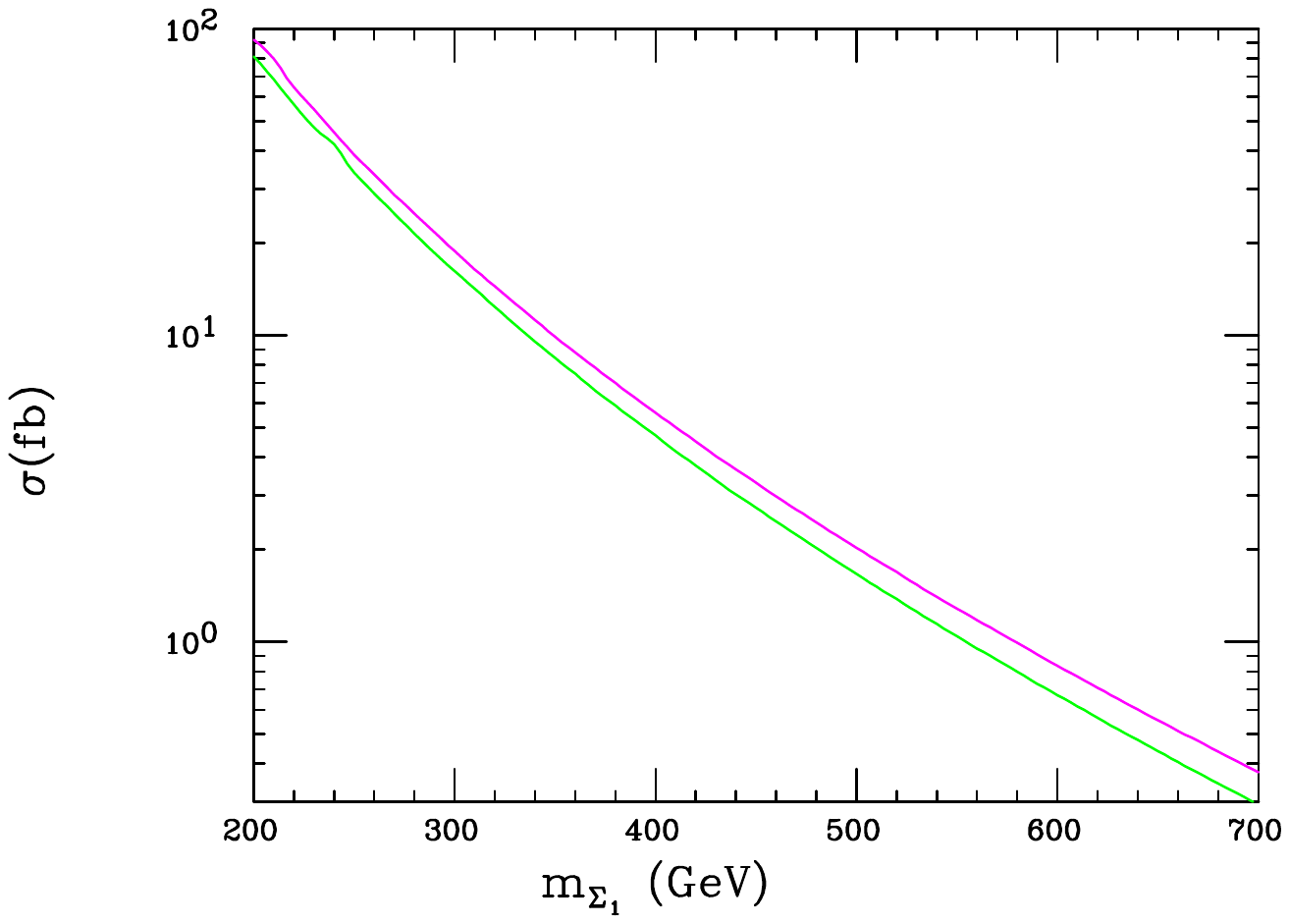}}
\vspace*{-1.3cm}
\caption{$\Sigma_1$ (or $\Sigma_2$) pair production cross section via the $q\bar q$ annihilation process as a function of its mass at the $\sqrt s=13$ (lower green) and 14 GeV (upper magenta) 
LHC, respectively. }
\label{fig2}
\end{figure}

Perhaps the process with the least sensitivity to the detailed nature of the PM scalar sector occurs for the pair production of the lightest of the charged scalars, $\Sigma_1$, via the process 
$\bar q q\to \gamma^*,Z^*\to \Sigma_1^+\Sigma_1^-$. As was noted above, the two neutral states, $h_D,{\cal H}$, are not accessible via $Z$ exchange at they both have $T_3=0$. 
The rate for $\Sigma_1$ pair production is seen to be {\it insensitive} to the value of $t$, \ie, the relative magnitudes of $s_\phi^2$ and $c_\phi^2$, depending only upon the value of 
$m_1$ and not on the masses of the other scalars. Note that although the values of $m_1$ have been chosen to be relatively small for our two benchmarks, a scan of the enlarged model parameter 
space allows $\Sigma_1$ masses possibly as large as roughly $\simeq 450-500$ GeV in some cases. Of course, once the $\Sigma_1$'s are produced on-shell, their decays {\it are} controlled by the 
value of $t$ as well as by the masses of the remaining scalars. Fig.~\ref{fig2} shows this production cross section at the $\sqrt s=13,14$ TeV LHC; of course this same result will hold in the case of 
$\Sigma_2$ pair production (simply with $m_1 \to m_2$ since the absolute values of the $\gamma,Z$ couplings for $\Sigma_{1,2}$ are identical) except that these states are significantly more 
massive so result in a much smaller production rate{\footnote {Since the $\gamma,Z$ couplings are diagonal as $\Sigma_{1,2}$ do not mix until terms of order $x_{t,s}^2$ are 
included, $\Sigma_1^\pm\Sigma_2^\mp$ production is essentially absent.}}.  

How will $\Sigma_1$ decay once it is produced? Although it is suppressed by $s_\phi^2$, we observe that the two-body 
decay mode, $\Sigma_1^+ \to W^+h_D$, is always kinematically open as $h_D$ is very light. In principle, there is a second possibility, $\Sigma_1 \to W{\cal H}^*, {\cal H}^*\to all$, but given the mass 
spectra of our generalized benchmark models this is (at least) a 
three-body decay and so the corresponding partial width is significantly suppressed even though it is instead proportional to the larger $c_\phi^2$ factor. This implies that the branching fraction 
for $\Sigma_1^+ \to W^+h_D$ is essentially unity. That being the case, $\Sigma_1$ pair production and subsequent decay will lead to the $W^+W^-+$MET final state in a manner qualitatively similar 
to charged wino pair production - apart from the specific spin-dependent differences in the production rate for a given parent mass and the angular distributions of the final state particles. 

The ATLAS Collaboration has performed several searches for this final state, depending upon the possible decay modes of the $W$'s, for the production and decay of a pair of charged winos, 
which is just the fermionic analog of the present process, and which are summarized (and combined) in Ref.~\cite{ATLAS:2024qxh}. Of course, although the threshold behavior of the cross  
section and the acceptances will be somewhat different for the chargino case due to the difference in particle spin, \etc, perhaps these analyses can give us a (very) crude handle on the present 
LHC mass reach sensitivity for $\Sigma_1$ by converting their chargino pair mass bounds, for massless LSPs, into cross section limits, then applying them to the case at hand. For example, the 
resulting $2\ell 0j~(1\ell)$ chargino bound, reinterpreted in the present case at face value, would exclude the range $m_1 < 230(280)$ GeV. Similarly, the analogous search in the all hadronic 
channel by ATLAS would exclude the mass window $330 < m_1< 420$ GeV, which is where are two benchmarks lie. The CMS Collaboration has performed a similar search in the case of the 
all hadronic final state\cite{CMS:2022sfi} and would correspondingly translate into a lower bound of $m_1>370$ GeV; their corresponding $2\ell 0j$ channel search would not exclude any of the 
present model parameter space\cite{CMS2}.  The big problem with employing this straightforward translational approach is not just the differences due to the angular distributions of the parent 
particle in each case but also their differing masses, corresponding to identical production cross sections,  which will influence the acceptances via the $p_T$'s of the leptons and/or 
jets from the $W$ that are needed to pass a given set of analysis cuts. These acceptances and efficiencies for a given analysis will certainly differ by $O(1)$ factors which will strongly influence any 
bound we quote. While it is clear that these analyses when properly applied to the current case will likely cut into our model parameter space somewhat, perhaps even showing sensitivities near 
our fixed benchmarks, a much more thorough analysis than that provided here is absolutely necessary before any conclusions can be drawn. This is especially true if the three analyses employed 
for the chargino searches are to be combined, as was done in Ref.~\cite{ATLAS:2024qxh}, so as to obtain a stronger sensitivity to the model space.

\begin{figure}[htbp] 
\centerline{\includegraphics[width=5.0in,angle=0]{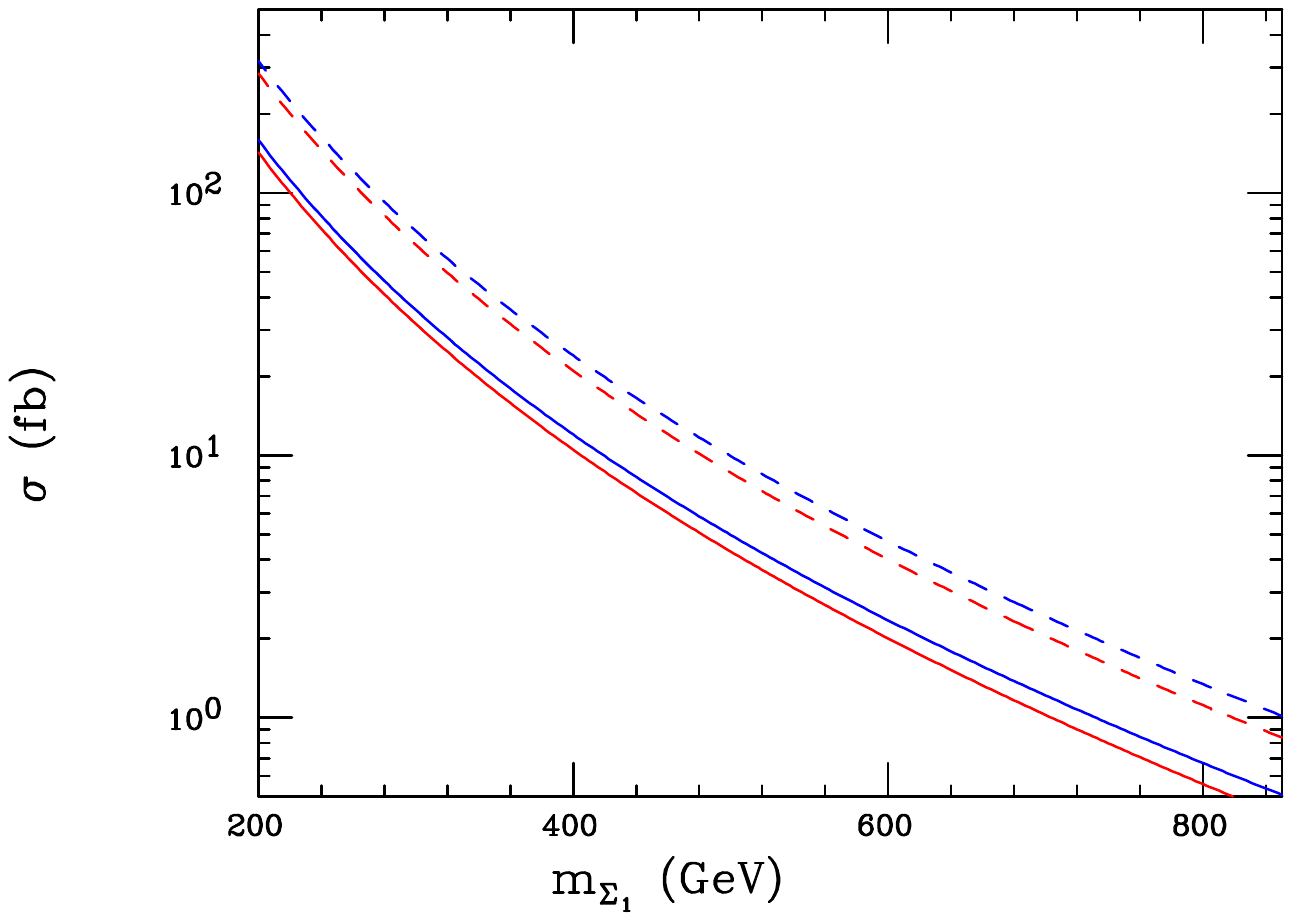}}
\vspace*{-1.3cm}
\caption{The $\Sigma_{1}h_D^\dagger+h.c.$ associated production cross section as a function of $m_{1,2}$ for the generalized benchmark models BM1 (solid) and BM2 (dashed) with the case of the 
$\sqrt s =13 (14)$ TeV LHC being the red lower (blue upper) member of each pair of curves, respectively, as described in the text. }
\label{fig3}
\end{figure}

Unless one of the two $W$'s is missed, \eg, in the $2\ell 0j$ analysis, however, this production process will not supply a large background to the $WV$-induced $W$+MET signal that we were 
looking for as evidence of possible dark moment $W$ couplings but the considerations going into the previous discussion above have pointed us in a new direction. Since we know that 
$\Sigma_1$ dominantly decays to $Wh_D$, the associated production process $\bar q q'\to W^* \to \Sigma_1 h_D^\dagger +h.c.$, followed by $\Sigma_1$ decay {\it does} lead to the $W+2h_D$, 
\ie,  the $W$+MET final state, albeit with a more model-dependent rate than simple $\Sigma_1$ pair production. Here it will be assumed that the $h_D$ is either long-lived or, more likely, will decay 
directly or indirectly into DM, thus appearing as MET. This is the simplest (at least partially visible) final state signature that arises for the production of the PM scalars in the present setup. The rate for 
this process will be sensitive not only to the value of $m_1$ but now also the value of $t$ which determines the size of the mixing angle suppression, $s_\phi^2$, that results from the same vertex as 
is responsible for $\Sigma_1$ decay, $W\Sigma_1h_D$. Fig.~\ref{fig3} shows the the production cross section for this process as a function of $m_1$ holding the value of $t$ fixed to those of our 
two benchmark models, but now generalized to the case where $m_1$ is treated as a free parameter. Here all the scalar mass ratios are still being held fixed to their benchmark values for 
demonstration purposes.  Also, the same cross section result will hold for the case of $\Sigma_2 h_D$ production except for the larger mass, \ie, $m_1 \to m_2$, as in the $\Sigma_1$ pair 
production case above. These cross sections are seen to be quite sizable for values of $m_1\lsim 400$ GeV or so, being roughly 2 orders of magnitude or more larger than those for $WV$ 
production before any further kinematic cuts are applied, clearly making the dark moment couplings of the $W$ even more difficult to access.

\begin{figure}[htbp]
\centerline{\includegraphics[width=5.0in,angle=0]{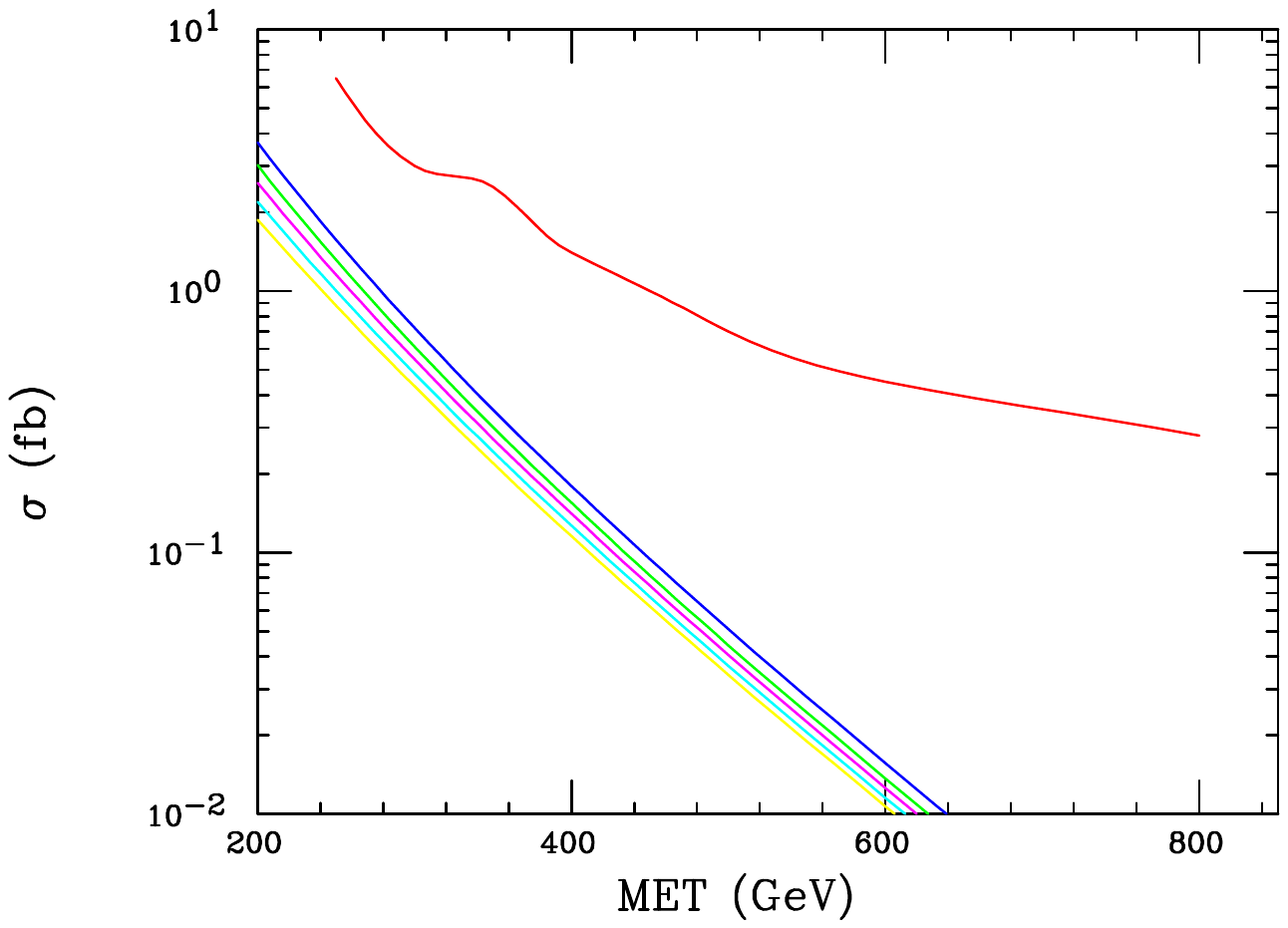}}
\vspace*{-0.9cm}
\centerline{\includegraphics[width=5.0in,angle=0]{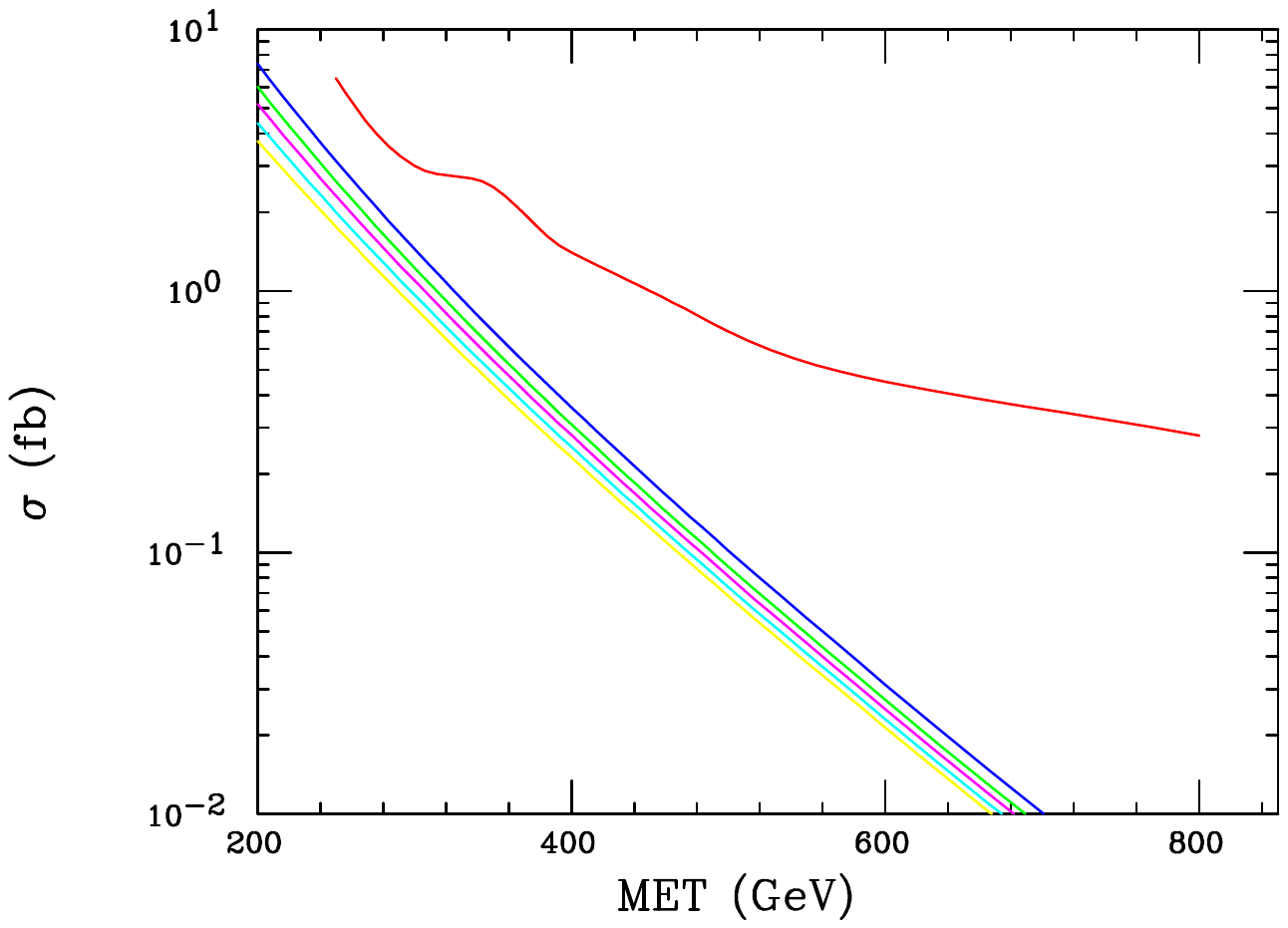}}
\vspace*{-1.3cm}
\caption{Cross sections, as functions of the cut on the missing transverse energy, for the $W^\pm (\to ~hadrons)$+MET final state arising from the $\Sigma_1h_D^\dagger+h.c.$ production process 
at the $\sqrt s=13$ TeV LHC for the generalized versions of the two benchmark models, (Top) BM1 and (Bottom) BM2, respectively. These are compared to the strictest upper bounds set by ATLAS  
from the high purity merged jet analysis as are given in Ref.\cite{ATLAS:2024rlu}, which roughly corresponds to the upper red curve in both panels. From top to bottom in each panel, the set of 
rapidly falling curves correspond to the choice of $m_1=200-400$ GeV, in steps of 50 GeV.}
\label{fig4}
\end{figure}

This final state, with the $W$ decaying hadronically, was employed in our earlier work\cite{Rizzo:2022jti} as a possible probe of the model parameter space and it is again useful here, not only in 
this regard, but also to show how much this {\it additional} new physics signal arising from this same model will swamp that coming from the previously considered $WV$ final state. Here we need 
to weight our previous cross sections by the hadronic branching fraction of the $W$\cite{ParticleDataGroup:2024cfk} as well as by the appropriate mixing factor. Since our earlier work, 
the LHC has accumulated roughly a factor of $\simeq 4$ times the integrated luminosity and the ATLAS Collaboration has improved upon their analysis that we had followed earlier, 
Ref.\cite{ATLAS:2018nda}, in their more recent work in Ref.\cite{ATLAS:2024rlu}, which we now employ. From the model side, we have now moved into the $t<1$ region implying that 
$s_\phi^2 <1/2$ which then suppresses the production of the final state of interest. Note that as long as we stay in this region, cross sections larger that those for BM2 by a factor $\simeq 1.9$ 
cannot be obtained. In particular, we see that BM1 and BM2 lead to cross sections which are now roughly a factor of $\simeq 4-7$ times smaller than those considered in our previous analysis. 
Following our earlier procedure, Fig.~\ref{fig4} displays the result of these considerations comparing the predictions of our two generalized benchmark models for values of $m_1$ lying in the 
range 200-400 GeV as a function of MET in comparison to the {\it tightest} constraint obtained in Ref.\cite{ATLAS:2024rlu} from the analysis where `high purity' jets from the $W$ decay are merged. 
Recall that as $m_1$ increases the model will more easily pass a 
given MET threshold but, at the same time, the overall cross section is also decreasing thus reducing the predicted $m_1$-dependent differences. This we find results in the rather steeply falling 
set of curves for both of the benchmarks in this Figure. Note that at lower values of MET, $\simeq 250-300$ GeV, the predictions of BM2 for smaller values of $m_1$ lie closest to the ATLAS 
exclusion curve - approaching within roughly a factor of $\simeq 2$. Thus with larger values of $t$ even smaller values of $m_1$ would certainly lie rather close to the present exclusion curve. 
The corresponding analysis by CMS\cite{CMS:2021far} is more difficult to interpret in a model independent fashion, but something we would want to employ to confirm these conclusions.  Clearly, 
further accumulation of integrated luminosity and analysis improvements will begin to probe closer to our overall model parameter space using this final state and a much more sophisticated 
analysis would certainly be warranted. We further note that a comparison of these results with those displayed in Fig.~\ref{fig0} shows that the direct production of the PM fields in this model by just this 
single mechanism will completely swamp any of the corresponding signals arising from either ordinary KM mixing or that from dark $W$ moments considered here by up to two orders of magnitude 
or more.

\begin{figure}[htbp] 
\centerline{\includegraphics[width=5.0in,angle=0]{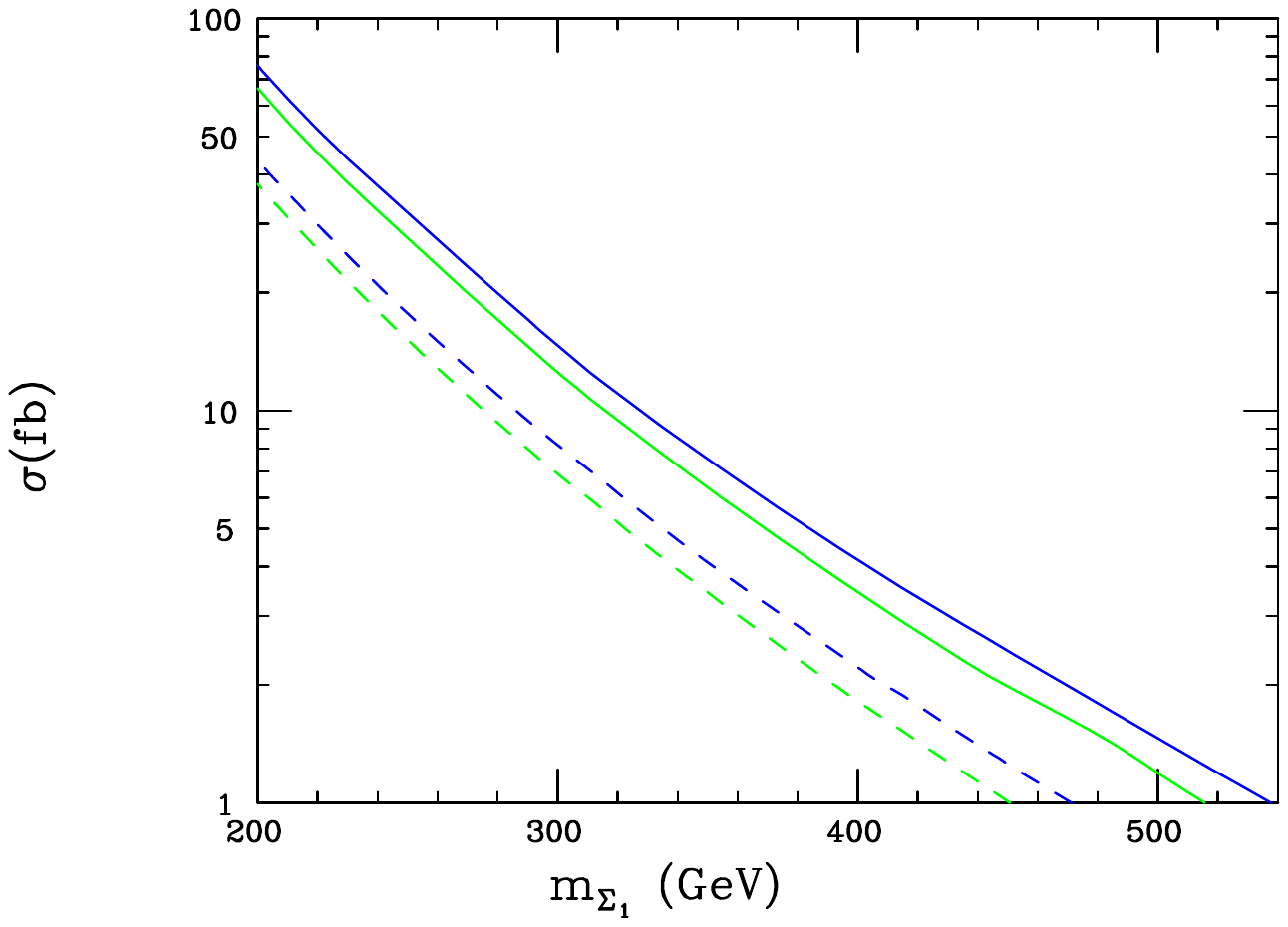}}
\vspace*{-0.9cm}
\centerline{\includegraphics[width=5.0in,angle=0]{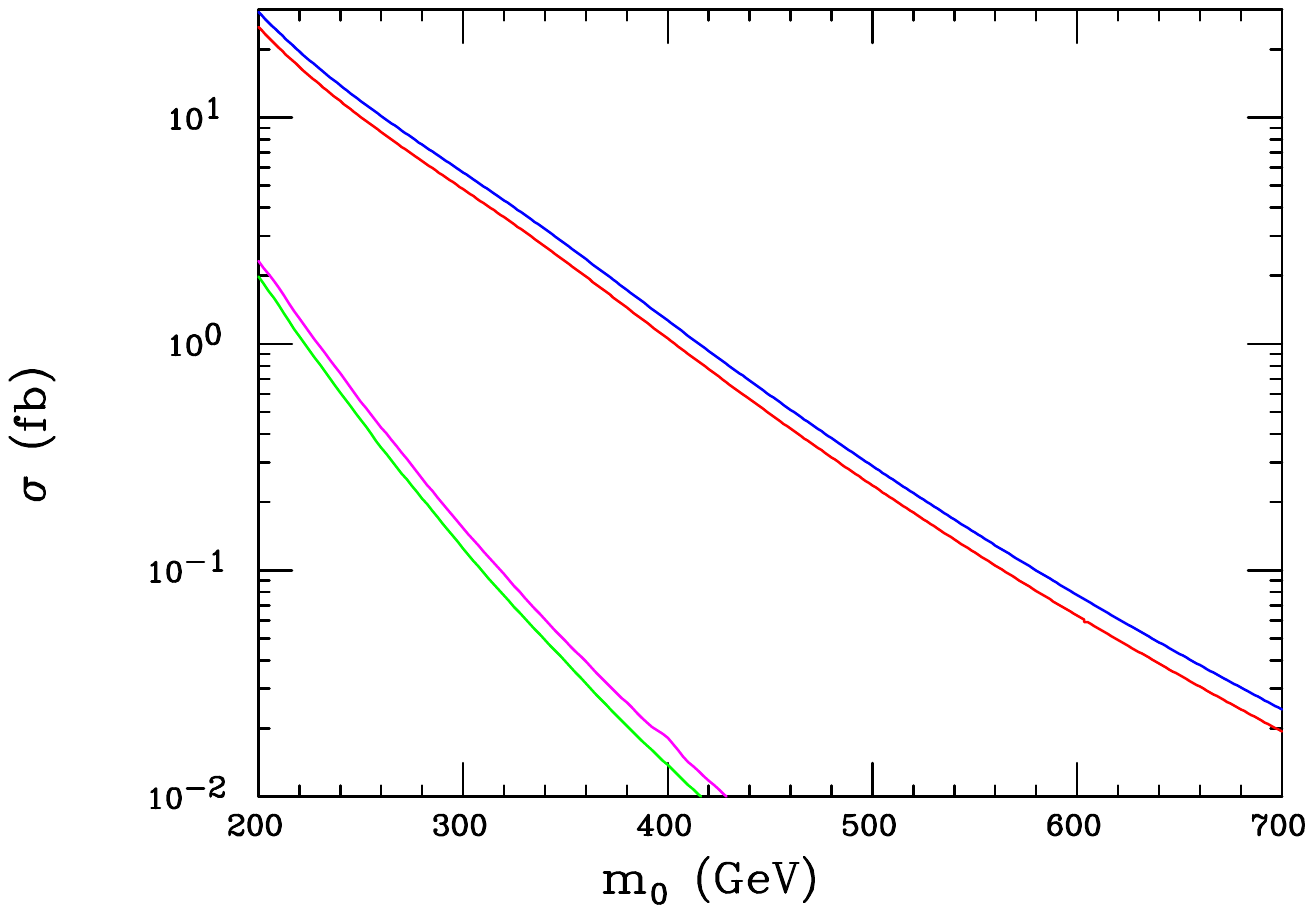}}
\vspace*{-1.3cm}
\caption{(Top) Associated production cross section for the $\Sigma_1{\cal H}+h.c.$ final state for the generalized benchmark models BM1 (solid) and BM2 (dashed) as a function of $m_1$. The 
lower green (upper blue) curve in each case corresponds to the result for the $\sqrt s=13(14)$ TeV LHC, respectively.  (Bottom) Cross section for the SM Higgs mediated associated production 
of $h_D{\cal H}^\dagger+h.c.$ (top pair of curves) and for the ${\cal H}$ pair production (lower pair of curves) final states, here as a functions of $m_0=m_{\cal H}$. For either process, the 
bottom (top) curve of the pair is for the $\sqrt s=13(14)$ TeV LHC, respectively. Here, for purposes of demonstration, we have assumed that quartic couplings satisfy $\lambda'',\lambda'''=1$. }
\label{fig5}
\end{figure}

So far, we have mainly concerned ourselves with the lighter PM states, $h_D,\Sigma_1$ (and, somewhat indirectly, also the heavier charged state, $\Sigma_2$ as far as production cross sections 
go). The neutral complex scalar state, ${\cal H}$, can also be produced in several ways leading to final states similar to those already discussed above. Overall, however, this complex neutral state 
is a bit less straightforward to produce and analyze. As in the case of virtual $W$ exchange leading to the $\Sigma_1 h_D^\dagger +h.c.$ final state, the corresponding mode, 
$\Sigma_1 {\cal H}^\dagger +h.c.$, is also potentially accessible but suffers from some phase space suppression which 
is not compensated for by the replacement of $s_\phi^2$ by the somewhat larger $c_\phi^2$ appearing in the expression for the cross section. The rate for this process for our two generalized 
benchmark models can be found in the upper panel of Fig.~\ref{fig5} where we see that these cross sections still have respectable magnitudes. Recall again that for these two generalized 
benchmarks the mass of ${\cal H}$ is completely determined by the choice of the value of the mass parameter $m_1$. 

Subsequent to ${\cal H}$ production, the two-body decay path ${\cal H}\to W^-\Sigma_1^+$ is generally kinematically allowed (as well as being allowed by $Q_D$ conservation, unlike 
the analogous, opposite-signed $W^+\Sigma_1^-$ mode) for almost all of both generalized benchmark parameters spaces. As discussed above, since subsequently $\Sigma_1^+\to W^+h_D$, 
this ${\cal H}$ decay mode essentially leads to the $W^+W^-$+MET final state with an appreciable but more model-dependent rate, as was also the case for $\Sigma_1$ pair production itself as 
we saw above, but with differing kinematics. We note, of course, that this is not the only 
possible decay path as can be seen from our earlier work\cite{Rizzo:2021lob} where the trilinear couplings between the SM Higgs and the PM scalars to leading order in $x_{t,s}$ (and so 
proportional to $v_d$) were briefly investigated. In the present notation, these interactions are seen to take the general form given by 
\begin{equation}
\begin{aligned}
\mathcal{L} \supset ~\frac{h_{SM} v_d}{2} &\bigg\{2(h_D^\dagger h_D) \left[\lambda'=~\frac{t^2}{2(1+t^2)}(\lambda_{4\Phi}+\lambda_5)+\frac{\kappa}{1+t^2}-\frac{\sqrt 2 ~\tilde \lambda ~t}{1+t^2}\right]  \\
&+(h_D^\dagger {\cal H}+h.c. ) \left[ \lambda''~=\frac{t}{1+t^2}(\lambda_{4\Phi}+\lambda_5-2\kappa)-\sqrt 2 ~\tilde \lambda ~\frac{1-t^2}{1+t^2}\right] \\
&+2({\cal H}^\dagger {\cal H}) \left[ \lambda'''~=\frac{1}{2(1+t^2)}(\lambda_{4\Phi}+\lambda_5)+\frac{\kappa t^2}{1+t^2}+\frac{\sqrt 2~ \tilde \lambda ~t}{1+t^2}\right]\\ 
&+ 2\lambda_{4\Phi} \Sigma_1^+\Sigma_1^- +2(\lambda_{4\Phi}+\lambda_5)\Sigma_2^+\Sigma_2^- \bigg\}.
\end{aligned} 
\end{equation}
The first term, proportional to $\lambda'$, leads to the invisible decay of the SM Higgs, while the second term, proportional to $\lambda''$, allows for the decay ${\cal H}\to h_Dh_{SM}$ which is 
of immediate interest to us. We note that this same term also will allow for $h_D^\dagger {\cal H}+h.c.$ production via off-shell $h_{SM}$ exchange. Similarly, the $\lambda'''$ term and those on the 
last line of this Equation can lead to ${\cal H}{\cal H}^\dagger$ as well as $\Sigma_i^+\Sigma_i^-$ pair-production, respectively, by the same virtual SM Higgs exchange mechanism. In the later 
cases, we already know from above that $\Sigma_i$ pair production is by far dominated by virtual $\gamma,Z$ exchange, as will be more fully seen below. The partial width for the decay 
${\cal H}\to h_Dh_{SM}$ is given by (taking $m_{h_D}\to 0$) 
\begin{equation}
\Gamma({\cal H}\to h_Dh_{SM})=\frac{(\lambda'')^2v_d^2}{64\pi m_{\cal H}}~(1-r_h)\,,
\end{equation}
while that for ${\cal H}\to W^-\Sigma_1^+$ is instead given by 
\begin{equation}
\Gamma({\cal H}\to W^-\Sigma_1^+)=c_\phi^2~\frac{G_Fm_{\cal H}^3}{2\sqrt 2 \pi}~\big[(1+r_1-r_W)^2-4r_1\big]^{1/2}~\Big[(1-r_1)^2-2r_W(1+r_1)+r_W^2\Big]\,,
\end{equation}
where we have defined the mass ratios $r_{W,1,h}=m_{W,1,h_{SM}}^2/m_{\cal H}^2$ in the expressions above. Neglecting any phase suppression for the moment, we see that the ratio of these 
two partial widths is just given by 
\begin{equation}
{\cal R}=\frac{c_\phi^2}{(\lambda'')^2}~\Big(\frac{2m_{\cal H}}{v_d}\Big)^4\,,
\end{equation}
which we see is $\sim 200$ when $\lambda'' \simeq 1$ for either benchmark model. From this we can conclude that, unless $\lambda''$ is very large or phase space is very significantly closed, 
then the $W^-\Sigma_1^+$ decay mode of the ${\cal H}$ will generally be the by far dominant one.  

Of course, one may ask just how large the $\lambda''$ (or $\lambda'''$) quartic may be given that we already know that $\lambda'$ must be quite small, \ie,  $|\lambda'|\lsim 6.5\cdot 10^{-3}$, to avoid 
the LHC bounds on the invisible width of the Higgs\cite{Invisible} - as all three of these quartics involve rather similar combinations of parameters. To get a very rough feel for this, we can perform a 
simple numerical exercise where we imagine that $\lambda'$ (for some reason) is {\it exactly} zero, which, though small, will certainly not be the case in reality. Then we can solve for the combination 
$\lambda_{4\Phi}+\lambda_5$ in terms of $\tilde \lambda, \kappa$ and $t$ and insert this result into the expression for 
$\lambda''$ and then express $\tilde \lambda$ in terms of $m_{1,2}$ using Eq.(5) above. For a given benchmark model, where $m_{1,2}$ and $t$ are fixed, we can then express $\lambda''$ solely in 
terms of the only remaining unknown variable, $\kappa$. Doing so, we then find the numerical results  
\begin{equation}
\lambda''=5.81-4.93\kappa~(10.00-3.19\kappa)\,,
\end{equation}
for the case of BM1(BM2), respectively. Clearly this implies that $\lambda''$ is O(1) or even larger unless $\kappa$ is itself appropriately tuned in a benchmark-dependent fashion. Analogously, for 
the case of the $\lambda'''$ quartic, we find a somewhat similar pair of results if we make the same set of numerical assumptions: 
\begin{equation}
\lambda'''=14.31-5.07\kappa~(15.92-1.54\kappa)\,,
\end{equation}
for the case of BM1(BM2). Note that with this limited parameter freedom it is not possible to make both $\lambda''$ and $\Lambda'''$ small simultaneously in either benchmark model case. 
It is clear from these brief considerations that both $\lambda''$ and $\lambda'''$ will very likely be (at least) of order unity. However, unless $\lambda''$ is non-perturbatively 
large, we would still expect the ${\cal H} \to W^-\Sigma_1^+$ decay mode to dominate. Repeating this short study for any small, realistic value of $|\lambda'|\lsim 6.5\cdot 10^{-3}$ leads to 
semiquantitatively very similar results and the same overall conclusions. Still, the fact that both of these couplings should generally be expected to be large makes us want to re-examine the 
production cross sections for both the $h_D^\dagger {\cal H}+h.c.$ and ${\cal H}^\dagger {\cal H}$ final states via an off-shell SM Higgs exchange with this possibility in mind.

The lower panel of Fig.~\ref{fig5} shows the associated production cross section for $h_D{\cal H}^\dagger+h.c.$ (upper pair of curves) as well as for ${\cal H}{\cal H}^\dagger$ pair production 
(lower pair of curves) via $gg\to h_{SM}^*$ exchange as a function of $m_0=m_{\cal H}$ at the $\sqrt s=13,14$ TeV LHC when both $\lambda''$ and $\lambda'''$ are taken equal to unity. We note 
that the lower set of curves would also be applicable to the cases of virtual $h_{SM}^*$ exchange production of the $\Sigma_i^\dagger \Sigma_i$ final states with appropriately scaled quartic factors. 
In such cases we find that indeed these cross sections are seen to be far smaller that those arising from the virtual $\gamma,Z$ exchange mechanism as we discussed above unless the relevant 
quartics were to be quite large $>>1$. While the $h_D{\cal H}^\dagger+h.c.$ mode is observed to have a reasonable rate, that for ${\cal H}{\cal H}^\dagger$ is significantly phase space 
suppressed {\it unless} the responsible quartic were again to be much larger than unity. From the decay discussions above, we see that $h_D{\cal H}^\dagger+h.c.$ production again leads to the  
$W^+W^-+$MET final state while ${\cal H}{\cal H}^\dagger$ production will instead lead to a $2W+2W^-$MET final state which would be rather challenging to reconstruct and analyze.

\section{Discussion and Conclusion}

The familiar vector boson/kinetic mixing portal scenario posits an interaction between the SM and dark sector fields that arises solely via the kinetic mixing of the SM and dark photons due 
to vacuum polarization-like loops of other new scalar and/or fermionic particles carrying both SM and dark charges, \ie, portal matter. However, depending upon the exact nature of these 
particles, \ie, their transformation properties under the SM gauge group, in earlier work it was shown that PM can also generate other loop graphs that lead to additional forms for the SM fermion-DP 
interaction. These are similar to the more familiar anomalous charge and magnetic moment form factors and can have important phenomenological implications. Such interactions could even 
dominate over those due to ordinary KM mixing in the determination of the DM relic density and at producing new signatures at both fixed target and collider experiments. However, these new 
interactions for the SM fermions were generally found to also require an extension of the usual $U(1)_D$ gauge group to some larger, non-abelian structure in which some of the SM fermions  
transform in a non-trivial manner.  One purpose of the present paper was to determine if such sizable new interactions could be also generated for the SM $W$ gauge boson via scalar PM loops 
and whether or not they can be lead to an enhanced rate for the $WV$ final state signal at the LHC, which appears as $W+$MET, that might then be visible above the known SM backgrounds. 
This analysis was performed by employing a previously introduced scalar PM model having an extended Higgs sector with an additional complex $SU(2)_L$ triplet as well as a complex singlet set 
of scalar fields, both of which carry non-zero dark charges beyond the usual SM Higgs doublet which has $Q_D=0$. However, it is to be noted that unlike in the case of the previously examined fermionic PM-induced dark moment type couplings, here no extension of the $U(1)_D$ dark gauge group was required for this goal to be achieved although it will be interesting to contemplate 
such gauge extensions in future study. In this setup the size of these induced dark moments were found to be relatively small. It seems quite likely that in the more complex non-abelian case, unless a 
clever means is found to avoid the degeneracy issues as encountered above, the anticipated result in such a setup would be similar in magnitude to those discussed here for PM scalars up to 
overall numerical factors. The question would then be whether or not such factors may be phenomenologically significant in raising the $WV$ cross section to the visible level and this possibility will 
be examined elsewhere. 

In this paper it was shown that the rate for the production of the $WV$ final state could be significantly enhanced over that predicted by the usual KM setup - by almost 2 orders of magnitude in 
some parameter space regions - via a dark triple gauge interaction for the $W$. Even so, it was found that the size of such couplings in the present scalar PM model remained too small to lead to 
a visible signal above the large SM backgrounds in this $W+$MET channel. While the $WV$ production rate might well become even more significantly enhanced at higher energy hadron colliders, 
such as the FCC-hh, it was noted that at some point it might be possible, and more advantageous, to instead search for the underlying physics, \ie, the direct production of the new scalar PM 
particles that populate the loop diagrams responsible for these dark anomalous 
moment couplings. In fact, in some regions of parameter space, it was shown that for direct production one doesn't need to wait for the FCC-hh as these PM scalars may already be accessible 
with significant rates at the HL-LHC. It was observed too that their production and decay necessarily would also further increase the apparent backgrounds for the, now more indirect, $WV$ final state 
production signature. However, it was also shown that these new 'background' sources can fortunately be turned into new physics signals for this underlying set of PM scalar fields. 

At leading order in the hierarchy of vev ratios for the mixings among the scalars, the PM fields do not couple to the SM fermions and the neutral PM scalars also do not couple to the SM $Z$, since 
they have $T_3=0$. Thus the production and subsequent cascade decay of these new particles all naturally lead to final states of the form $nW+$MET 
with $n=1-4$ depending upon the particular PM scalars position in the mass spectrum. Of these the $W^\pm+$MET and the $W^+W^-+$MET final states are of particular interest to us as searches 
have already been performed in these two channels as part of the general SUSY and dark sector programs of both ATLAS and CMS. In the discussion above we have seen that ($i$) the ATLAS 
$W^\pm(\to hadrons)+$MET, high purity merged jets search limit comes quite close to the BM2 benchmark model predictions for $\Sigma_1h_D$ production for lighter $\Sigma_1$ masses. 
Although a more detailed 
analysis of this signature is clearly warranted for this model, it seems more than likely the HL-LHC will be able to probe a reasonable fraction of the relevant model parameter space. ($ii$) Similarly, 
the corresponding searches by ATLAS in the all hadronic, $2\ell 0j$ and $1\ell$ channels targeting pairs of winos decaying to $W$'s plus the LSP were also shown to have some significant sensitivity 
to the process of $\Sigma_1$ pair production and subsequent decays. However, the differing acceptances and analysis efficiencies due to the distinction in the parent particles spin and cross section 
for a fixed parent particle mass value, however, makes any direct conclusions difficult without a realistic detector-level simulation study. Nonetheless, it is more than likely that the HL-LHC will also be 
able to probe a reasonable fraction of the parameter space of this scalar PM model using this final state as well.

Hopefully, some signals from the dark sector will soon be observed in the laboratory.

\section*{Acknowledgements}
The author would like to particularly thank J.L. Hewett for both hospitality and  wide ranging discussions. This work was supported by the Department of Energy, Contract DE-AC02-76SF00515.




\begin{thebibliography}{99}



\bibitem{Planck:2018vyg}
N.~Aghanim \textit{et al.} [Planck],
Astron. Astrophys. \textbf{641}, A6 (2020)
[erratum: Astron. Astrophys. \textbf{652}, C4 (2021)]
[arXiv:1807.06209 [astro-ph.CO]].


\bibitem{KM}
  B.~Holdom,
  Phys.\ Lett.\  {\bf 166B}, 196 (1986) and
  Phys.\ Lett.\ B {\bf 178}, 65 (1986); 
  K.~R.~Dienes, C.~F.~Kolda and J.~March-Russell,
  Nucl.\ Phys.\ B {\bf 492}, 104 (1997)
  [hep-ph/9610479];
  F.~Del Aguila,
  Acta Phys.\ Polon.\ B {\bf 25}, 1317 (1994)
  [hep-ph/9404323];
  K.~S.~Babu, C.~F.~Kolda and J.~March-Russell,
  Phys.\ Rev.\ D {\bf 54}, 4635 (1996)
  [hep-ph/9603212];
  T.~G.~Rizzo,
  Phys.\ Rev.\ D {\bf 59}, 015020 (1998)
  [hep-ph/9806397].

\bibitem{vectorportal} 
 There has been a huge amount of historical work on this subject; see, for example, 
  D.~Feldman, B.~Kors and P.~Nath,
  Phys.\ Rev.\ D {\bf 75}, 023503 (2007)
  [hep-ph/0610133];
  D.~Feldman, Z.~Liu and P.~Nath,
  Phys.\ Rev.\ D {\bf 75}, 115001 (2007)
  [hep-ph/0702123 [HEP-PH]].;
  M.~Pospelov, A.~Ritz and M.~B.~Voloshin,
  Phys.\ Lett.\ B {\bf 662}, 53 (2008)
  [arXiv:0711.4866 [hep-ph]];
  M.~Pospelov,
  Phys.\ Rev.\ D {\bf 80}, 095002 (2009)
  [arXiv:0811.1030 [hep-ph]]; 
  H.~Davoudiasl, H.~S.~Lee and W.~J.~Marciano,
  Phys.\ Rev.\ Lett.\  {\bf 109}, 031802 (2012)
  [arXiv:1205.2709 [hep-ph]] and 
  Phys.\ Rev.\ D {\bf 85}, 115019 (2012)
  [arXiv:1203.2947 [hep-ph]];
  R.~Essig {\it et al.},
  arXiv:1311.0029 [hep-ph];
  E.~Izaguirre, G.~Krnjaic, P.~Schuster and N.~Toro,
  Phys.\ Rev.\ Lett.\  {\bf 115}, no. 25, 251301 (2015)
  [arXiv:1505.00011 [hep-ph]];
  M.~Khlopov,
  Int.\ J.\ Mod.\ Phys.\ A {\bf 28}, 1330042 (2013)
  [arXiv:1311.2468 [astro-ph.CO]];
 For a general overview and introduction to this framework, see  
  D.~Curtin, R.~Essig, S.~Gori and J.~Shelton,
  JHEP {\bf 1502}, 157 (2015)
  [arXiv:1412.0018 [hep-ph]].

\bibitem{Gherghetta:2019coi}
T.~Gherghetta, J.~Kersten, K.~Olive and M.~Pospelov,
Phys. Rev. D \textbf{100}, no.9, 095001 (2019)
[arXiv:1909.00696 [hep-ph]].

 
\bibitem{Fabbrichesi:2020wbt}
M.~Fabbrichesi, E.~Gabrielli and G.~Lanfranchi,
[arXiv:2005.01515 [hep-ph]].

\bibitem{Graham:2021ggy}
M.~Graham, C.~Hearty and M.~Williams,
[arXiv:2104.10280 [hep-ph]].

\bibitem{Barducci:2021egn}
D.~Barducci, E.~Bertuzzo, G.~Grilli di Cortona and G.~M.~Salla,
JHEP \textbf{12}, 081 (2021)
[arXiv:2109.04852 [hep-ph]].

\bibitem{Alonso-Gonzalez:2025xqg}
D.~Alonso-Gonz{\'a}lez, D.~Cerde{\~n}o, P.~Foldenauer and J.~M.~No,
[arXiv:2507.11376 [hep-ph]].

\bibitem{Jorge:2026bbs}
A.~W.~R.~Jorge, L.~Sagunski, G.~W.~Yuan, T.~Song and E.~Bratkovskaya,
[arXiv:2601.15066 [hep-ph]].

\bibitem{Caputo:2026pdw}
A.~Caputo and R.~Essig,
[arXiv:2603.08430 [hep-ph]].

\bibitem{Li:2024wqj}
S.~Li, J.~M.~Yang, M.~Zhang and R.~Zhu,
[arXiv:2405.18226 [hep-ph]].


\bibitem{Rizzo:2018vlb}
T.~G.~Rizzo,
Phys. Rev. D \textbf{99}, no.11, 115024 (2019)
[arXiv:1810.07531 [hep-ph]].

\bibitem{Rueter:2019wdf}
T.~D.~Rueter and T.~G.~Rizzo,
Phys. Rev. D \textbf{101}, no.1, 015014 (2020)
[arXiv:1909.09160 [hep-ph]].

\bibitem{Kim:2019oyh}
J.~H.~Kim, S.~D.~Lane, H.~S.~Lee, I.~M.~Lewis and M.~Sullivan,
Phys. Rev. D \textbf{101}, no.3, 035041 (2020)
[arXiv:1904.05893 [hep-ph]].

\bibitem{Rueter:2020qhf}
T.~D.~Rueter and T.~G.~Rizzo,
[arXiv:2011.03529 [hep-ph]].

\bibitem{Wojcik:2020wgm}
G.~N.~Wojcik and T.~G.~Rizzo,
Phys. Rev. D \textbf{105}, no.1, 015032 (2022)
[arXiv:2012.05406 [hep-ph]].

\bibitem{Rizzo:2021lob}
T.~G.~Rizzo,
JHEP \textbf{11}, 035 (2021)
[arXiv:2106.11150 [hep-ph]].

\bibitem{Rizzo:2022qan}
T.~G.~Rizzo,
[arXiv:2202.02222 [hep-ph]].

\bibitem{Wojcik:2022rtk}
G.~N.~Wojcik,
[arXiv:2205.11545 [hep-ph]].

\bibitem{Rizzo:2022jti}
T.~G.~Rizzo,
Phys. Rev. D \textbf{106}, no.3, 035024 (2022)
[arXiv:2206.09814 [hep-ph]].

\bibitem{Rizzo:2022lpm}
T.~G.~Rizzo,
Phys. Rev. D \textbf{106}, no.9, 095024 (2022)
[arXiv:2209.00688 [hep-ph]].

\bibitem{Carvunis:2022yur}
A.~Carvunis, N.~McGinnis and D.~E.~Morrissey,
[arXiv:2209.14305 [hep-ph]].

\bibitem{Verma:2022nyd}
S.~Verma, S.~Biswas, A.~Chatterjee and J.~Ganguly,
[arXiv:2209.13888 [hep-ph]].

\bibitem{Rizzo:2023qbj}
T.~G.~Rizzo,
Phys. Rev. D \textbf{107}, no.9, 095014 (2023)
[arXiv:2302.12698 [hep-ph]].

\bibitem{Bauer:2022nwt}
M.~Bauer and P.~Foldenauer,
Phys. Rev. Lett. \textbf{129}, no.17, 171801 (2022)
[arXiv:2207.00023 [hep-ph]].


\bibitem{Wojcik:2022woa}
G.~N.~Wojcik, L.~L.~Everett, S.~T.~Eu and R.~Ximenes,
Phys. Lett. B \textbf{841}, 137931 (2023)
[arXiv:2211.09918 [hep-ph]].

\bibitem{Wojcik:2023ggt}
G.~N.~Wojcik, L.~L.~Everett, S.~T.~Eu and R.~Ximenes,
Phys. Rev. D \textbf{108}, no.5, 055033 (2023)
[arXiv:2303.12983 [hep-ph]].


\bibitem{Rizzo:2023kvy}
T.~G.~Rizzo,
Phys. Rev. D \textbf{108}, no.5, 055021 (2023)
[arXiv:2307.08508 [hep-ph]].

\bibitem{Rizzo:2023djp}
T.~G.~Rizzo,
Phys. Rev. D \textbf{109}, no.5, 055039 (2024)
[arXiv:2312.00226 [hep-ph]].

\bibitem{Rizzo:2024bhn}
T.~G.~Rizzo,
Phys. Rev. D \textbf{110}, no.7, 075037 (2024)
[arXiv:2408.01296 [hep-ph]].

\bibitem{Ardu:2024bxg}
M.~Ardu, M.~H.~Rahat, N.~Valori and O.~Vives,
[arXiv:2407.21100 [hep-ph]].

\bibitem{Rizzo:2024kzu}
T.~G.~Rizzo,
Phys. Rev. D \textbf{111}, no.7, 075018 (2025)
[arXiv:2412.17174 [hep-ph]].

\bibitem{Rizzo:2025tap}
T.~G.~Rizzo,
Phys. Rev. D \textbf{112}, no.1, 015029 (2025)
doi:10.1103/h81b-ydmw
[arXiv:2505.04474 [hep-ph]].

\bibitem{Tewary:2025vij}
K.~Tewary, S.~Biswas and S.~Verma,
[arXiv:2511.00578 [hep-ph]].

\bibitem{Rizzo:2025fsy}
T.~G.~Rizzo,
[arXiv:2511.13632 [hep-ph]].


\bibitem{CarcamoHernandez:2023wzf}
For a recent review of vector-like fermions, see 
A.~E.~C\'arcamo Hern\'andez, K.~Kowalska, H.~Lee and D.~Rizzo,
[arXiv:2309.13968 [hep-ph]].

\bibitem{CMS:2024bni}
A.~Hayrapetyan \textit{et al.} [CMS],
[arXiv:2405.17605 [hep-ex]].

\bibitem{Alves:2023ufm}
J.~M.~Alves, G.~C.~Branco, A.~L.~Cherchiglia, C.~C.~Nishi, J.~T.~Penedo, P.~M.~F.~Pereira, M.~N.~Rebelo and J.~I.~Silva-Marcos,
Phys. Rept. \textbf{1057}, 1-69 (2024)
[arXiv:2304.10561 [hep-ph]].

\bibitem{Banerjee:2024zvg}
A.~Banerjee, E.~Bergeaas Kuutmann, V.~Ellajosyula, R.~Enberg, G.~Ferretti and L.~Panizzi,
[arXiv:2406.09193 [hep-ph]].

\bibitem{Guedes:2021oqx}
G.~Guedes and J.~Santiago,
JHEP \textbf{01}, 111 (2022)
[arXiv:2107.03429 [hep-ph]].

\bibitem{Adhikary:2024esf}
A.~Adhikary, M.~Olechowski, J.~Rosiek and M.~Ryczkowski,
[arXiv:2406.16050 [hep-ph]].

\bibitem{Benbrik:2024fku}
R.~Benbrik, M.~Boukidi, M.~Ech-chaouy, S.~Moretti, K.~Salime and Q.~S.~Yan,
[arXiv:2412.01761 [hep-ph]].

\bibitem{Albergaria:2024pji}
F.~Albergaria, J.~F.~Bastos, B.~Belfatto, G.~C.~Branco, J.~T.~Penedo, A.~Rodr\'\i{}guez-S\'anchez and J.~I.~Silva-Marcos,
[arXiv:2412.21201 [hep-ph]].

\bibitem{Chen:2017hak}
C.~Y.~Chen, S.~Dawson and E.~Furlan,
Phys. Rev. D \textbf{96}, no.1, 015006 (2017)
[arXiv:1703.06134 [hep-ph]].

\bibitem{Biekotter:2016kgi}
A.~Biek{\"o}tter, J.~L.~Hewett, J.~S.~Kim, M.~Kr{\"a}mer, T.~G.~Rizzo, K.~Rolbiecki, J.~Tattersall and T.~Weber,
Int. J. Mod. Phys. A \textbf{32}, no.05, 1750032 (2017)
[arXiv:1608.01312 [hep-ph]].


\bibitem{Arcadi:2017kky} 
 G.~Arcadi, M.~Dutra, P.~Ghosh, M.~Lindner, Y.~Mambrini, M.~Pierre, S.~Profumo and F.~S.~Queiroz, 
Eur. Phys. J. C \textbf{78}, no.3, 203 (2018)
[arXiv:1703.07364 [hep-ph]].
  
\bibitem{Roszkowski:2017nbc}
L.~Roszkowski, E.~M.~Sessolo and S.~Trojanowski,
Rept. Prog. Phys. \textbf{81}, no.6, 066201 (2018)
[arXiv:1707.06277 [hep-ph]].

\bibitem{Arcadi:2024ukq}
G.~Arcadi, D.~Cabo-Almeida, M.~Dutra, P.~Ghosh, M.~Lindner, Y.~Mambrini, J.~P.~Neto, M.~Pierre, S.~Profumo and F.~S.~Queiroz,
[arXiv:2403.15860 [hep-ph]].
 

\bibitem{anom}
There is, of course, a huge set of literature on this general subject, reaching back over many decades both in general and in specific BSM scenarios; for a highly shortened sample of these 
see, for example,  
W.~A.~Bardeen, R.~Gastmans and B.~E.~Lautrup,
Nucl. Phys. B \textbf{46}, 319-331 (1972);
K.~J.~Kim and Y.~S.~Tsai,
Phys. Rev. D \textbf{7}, 3710 (1973);
K.~O.~Mikaelian, M.~A.~Samuel and D.~Sahdev,
Phys. Rev. Lett. \textbf{43}, 746 (1979);
R.~W.~Brown, D.~Sahdev and K.~O.~Mikaelian,
Phys. Rev. D \textbf{20}, 1164 (1979);
K.~Hagiwara, R.~D.~Peccei, D.~Zeppenfeld and K.~Hikasa,
Nucl. Phys. B \textbf{282}, 253-307 (1987);
U.~Baur and D.~Zeppenfeld,
Nucl. Phys. B \textbf{308}, 127-148 (1988);
G.~Couture, J.~N.~Ng, J.~L.~Hewett and T.~G.~Rizzo,
Phys. Rev. D \textbf{38}, 860 (1988);
G.~Couture, J.~N.~Ng, J.~L.~Hewett and T.~G.~Rizzo,
Phys. Rev. D \textbf{36}, 859 (1987);
T.~G.~Rizzo,
Phys. Rev. D \textbf{46}, 3894-3902 (1992);
A.~B.~Lahanas and V.~C.~Spanos,
Phys. Lett. B \textbf{334}, 378-390 (1994);
A.~Culatti,
Z. Phys. C \textbf{65}, 537-544 (1995);
F.~Larios, J.~A.~Leyva and R.~Martinez,
Phys. Rev. D \textbf{53}, 6686-6688 (1996);
A.~Moyotl and G.~Tavares-Velasco,
J. Phys. G \textbf{37}, 105012 (2010);
A.~Flores-Tlalpa, J.~Montano, H.~Novales-Sanchez, F.~Ramirez-Zavaleta and J.~J.~Toscano,
Phys. Rev. D \textbf{83}, 016011 (2011);
M.~A.~Arroyo-Ure{\~n}a, G.~Hern{\'a}ndez-Tom{\'e} and G.~Tavares-Velasco,
Phys. Rev. D \textbf{94}, no.9, 095006 (2016);
J.~Montano, G.~Tavares-Velasco, J.~J.~Toscano and F.~Ramirez-Zavaleta,
Phys. Rev. D \textbf{72}, 055023 (2005)
[arXiv:hep-ph/0508166 [hep-ph]].



\bibitem{CDF:2022hxs}
T.~Aaltonen \textit{et al.} [CDF],
Science \textbf{376}, no.6589, 170-176 (2022)

\bibitem{Peskin:1991sw}
M.~E.~Peskin and T.~Takeuchi,
Phys. Rev. D \textbf{46}, 381-409 (1992)

\bibitem{Peskin:1990zt}
M.~E.~Peskin and T.~Takeuchi,
Phys. Rev. Lett. \textbf{65}, 964-967 (1990)

\bibitem{Bozzi:2024lox}
For a recent review, see G.~Bozzi,
PoS \textbf{LHCP2024}, 051 (2025).


\bibitem{GBET}
  M.~S.~Chanowitz and M.~K.~Gaillard,
  Nucl.\ Phys.\ B {\bf 261}, 379 (1985);
  B.~W.~Lee, C.~Quigg and H.~B.~Thacker,
  Phys.\ Rev.\ D {\bf 16}, 1519 (1977);
  J.~M.~Cornwall, D.~N.~Levin and G.~Tiktopoulos,
  Phys.\ Rev.\ D {\bf 10}, 1145 (1974)
  Erratum: [Phys.\ Rev.\ D {\bf 11}, 972 (1975)];
  G.~J.~Gounaris, R.~Kogerler and H.~Neufeld,
  Phys.\ Rev.\ D {\bf 34}, 3257 (1986).


\bibitem{Invisible}
 ATLAS Collaboration, 
``Combination of searches for invisible Higgs boson decays with the ATLAS experiment,''
ATLAS-CONF-2020-052;
G.~Aad \textit{et al.} [ATLAS],
Phys. Lett. B \textbf{842}, 137963 (2023)
[arXiv:2301.10731 [hep-ex]];
V.~Milosevic [CMS],
PoS \textbf{EPS-HEP2021}, 602 (2022)


\bibitem{argyres}
E.~N.~Argyres and C.~G.~Papadopoulos,
Phys. Lett. B \textbf{263}, 298-304 (1991);
E.~N.~Argyres, A.~B.~Lahanas, C.~G.~Papadopoulos and V.~C.~Spanos,
Phys. Lett. B \textbf{383}, 63-77 (1996).

\bibitem{ParticleDataGroup:2024cfk}
S.~Navas \textit{et al.} [Particle Data Group],
Phys. Rev. D \textbf{110}, no.3, 030001 (2024)


\bibitem{Eichten:1984eu}
E.~Eichten, I.~Hinchliffe, K.~D.~Lane and C.~Quigg,
Rev. Mod. Phys. \textbf{56}, 579-707 (1984).

\bibitem{Kidonakis:2025jve}
N.~Kidonakis and A.~Tonero,
Eur. Phys. J. C \textbf{85}, no.11, 1270 (2025)
[arXiv:2506.01590 [hep-ph]].

\bibitem{ATLAS:2024qxh}
G.~Aad \textit{et al.} [ATLAS],
Phys. Rev. Lett. \textbf{133}, no.3, 031802 (2024)
[arXiv:2402.08347 [hep-ex]].

\bibitem{CMS:2022sfi}
A.~Tumasyan \textit{et al.} [CMS],
Phys. Lett. B \textbf{842}, 137460 (2023)
[arXiv:2205.09597 [hep-ex]].

\bibitem{CMS2}
CMS Physics Analyis Summary, CMS PAS SUS-23-002 (2024).


\bibitem{ATLAS:2018nda}
M.~Aaboud \textit{et al.} [ATLAS],
JHEP \textbf{10}, 180 (2018)
[arXiv:1807.11471 [hep-ex]].


\bibitem{ATLAS:2024rlu}
G.~Aad \textit{et al.} [ATLAS],
JHEP \textbf{11}, 126 (2024)
[arXiv:2406.01272 [hep-ex]].

\bibitem{CMS:2021far}
A.~Tumasyan \textit{et al.} [CMS],
JHEP \textbf{11}, 153 (2021)
[arXiv:2107.13021 [hep-ex]]


\bibitem{ATLAS:2024kpy}
G.~Aad \textit{et al.} [ATLAS],
Eur. Phys. J. C \textbf{84}, no.10, 1102 (2024)
[arXiv:2404.15930 [hep-ex]].










\end{thebibliography}
\end{document}